\documentstyle[aps]{revtex}

\def\ap{\alpha^{\prime}}
\def\qh{q^{1/2}}

\def\li{{{i}\over{l}}}
\def\lh{{{il}\over{2}}}
\def\upz{uz/\pi}

\def\uar{-{{iu}\over{\pi}} }

\def\ctu{{\rm Cos}(-2iu )}

\def\cul{{\rm Cos}(2ul)}
\def\culh{{\rm Cos}(ul)}

\def\su{{\rm Sin}(-iu)}
\def\apfour{{\textstyle{{{\alpha^{\prime}}\over{4}}}}}
\def\aptwo{{\textstyle{{{\alpha^{\prime}}\over{2}}}}}
\def\evac{{\textstyle{ - {{1}\over{24}} + {{\beta^2}\over{8}} }}}
\def\twofour{{\textstyle{ - {{1}\over{24}} }}}

\def\mmi{\half (1-\beta)}
\def\mpl{\half(1+\beta)}
\def\ssa{e^{-\pi i \alpha}}
\def\ssb{e^{-\pi i \beta}}
\def\L{{\cal L}}
\def\D{{\cal D}}
\def\Le{{\cal L_{E}}}
\def\A{{\cal A}}

\def\C{C^{\beta}_{\alpha}}

\def\2p{2\pi\alpha^{\prime}}
\def\sqap{{\textstyle{{\sqrt{ {{\alpha^{\prime}}\over{2}} }}}}}

\def\8p{8\pi^2\alpha^{\prime}}
\def\rmin{r_{\rm min.}}
\def\d0{{\rm D0brane}}
\def\delv{\delta V}

\def\delx{\delta X^{\mu}}
\def\delchi{\delta \chi_m}
\def\delchino{\delta \chi}
\def\delchinob{\delta{\bar \chi}}

\def\delpsi{\delta \psi}
\def\delpsib{\delta{\bar\psi}}
\def\delzeta{\delta \zeta}
\def\delxi{\delta \xi}
\def\delxib{\delta {\bar \xi}}
\def\ds{d^2 \sigma}
\def\sqg{{\sqrt{g}}}

\def\fidg{{\textstyle{{\sqrt{{\hat g}}}}}}
\def\area{\int d^2 \sigma \fidg}

\def\ponehalf{(p+1)/2} 
\def\phalf{p/2}

\def\eab{\epsilon^{ab}}
\def\gmn{g^{mn}}
\def\ema{e^a_m}
\def\emb{e^b_m}
\def\enb{e^b_n}
\def\ena{e_n^a}
\def\Xmu{X^{\mu}}

\def\gmu{\gamma^{\mu}}
\def\gnu{\gamma^{\nu}}

\def\ga{\gamma^{a}}
\def\gb{\gamma^{b}}

\def\gm{\gamma^{m}}
\def\gn{\gamma^{n}}
\def\pa{\partial_a}
\def\pb{\partial_b}
\def\pn{\partial_n}
\def\pzero{\partial_0}
\def\pone{\partial_1}
\def\ptwo{\partial_2}
\def\stwo{\sigma^2}
\def\sone{\sigma^1}

\def\gzero{\gamma^{0}}
\def\gone{\gamma^{1}}
\def\gtwo{\gamma^{2}}
\def\gfive{\gamma^{5}}
\def\joz{{\cal J}^{10}}
\def\jmunu{{\cal J}^{\mu\nu}}
\def\wmunu{\omega_{\mu\nu}}
\def\dl{D(\Lambda)}
\def\dli{D(\Lambda)^{-1}}
\def\psipl{\psi^+}
\def\psimi{\psi^-}

\def\ximi{\xi^-}
\def\chipl{\chi^+}

\def\psibmu{{\bar{\psi}}^{\mu}}

\def\psib{{\bar{\psi}}}

\def\xib{{\bar{\xi}}}
\def\chib{{\bar{\chi}}}
\def\zetab{{\bar{\zeta}}}
\def\halft{{{1}\over{2\pi\alpha'}}}
\def\quaft{{{1}\over{4\pi\alpha'}}}
\def\half{{\textstyle{1\over2}}}

\def\itwo{{\textstyle{i\over2}}}
\def\ifour{{\textstyle{i\over4}}}

\begin{document}
\draft
\preprint{PSU-TH-225}
\title{
Supersymmetric Pair Correlation Function of Wilson Loops} 
\author{Shyamoli Chaudhuri and Eric G.\ Novak 
\footnote{shyamoli@phys.psu.edu,novak@phys.psu.edu}}
\address{Physics Department \\
Penn State University \\ 
University Park, PA 16802}
\date{\today}
\maketitle
\begin{abstract}
We give a path integral derivation of the annulus diagram in a 
supersymmetric theory of open and closed strings with Dbranes. We 
compute the pair correlation function of Wilson loops in the 
generic weakly coupled supersymmetric flat spacetime background 
with Dbranes. We obtain a $-u^4/r^9$ potential between heavy 
nonrelativistic sources in a supersymmetric gauge theory at 
short distances.
\end{abstract}
\pacs{PACS numbers: 11.25.-w, 12.38.Aw, 31.15.Kb}

\section{Introduction}
\label{sec:intro}

Quantitative universal predictions for the low energy limit of String/M Theory
that are independent of specific backgrounds or compactifications are hard to
come by. In this paper, we take a step in the direction of quantitative 
universality, extracting a universal numerical result for the pair potential at short distances
between heavy nonrelativistic sources in a supersymmetric gauge theory on 
some generic D-manifold \cite{gimpol,polchinskibook} 
background of the type I or type II 
string theories. The result applies irrespective of whether the manifold, $\cal D$,
is the worldvolume of a higher dimensional Dbrane, the intersection of multiple 
Dbranes, or the bulk transverse space orthogonal to some configuration of 
branes.

Our result is obtained from a path integral prescription for the pair 
correlation 
function of Wilson loops living in some D-manifold in a weakly coupled 
background
of the type II theory, based on the earlier works 
\cite{alvarez,poltorus,cohen,dhokercomp,ccn}. 
We give a boundary reparameterization invariant computation of the 
supersymmetric pair correlation function of Wilson loops in the 
open and closed fermionic string theory 
\cite{dzbdh,howe,polyakov,durhuus,polcai} with Dbranes \cite{dbrane,polchinskibook}. 
The normalization of the one-loop string vacuum amplitude in such a background 
can be determined from first principles following a classic method due to
Polchinski \cite{poltorus}. However, it will be apparent from 
our result that the prediction of a short distance potential originating in 
fluctuations in the vacuum energy density is likely to hold in the broader 
context of the generic background of String/M theory.
Our prescription for determining the phase ambiguities in the fermionic string path 
integral is derived from the imposition of infrared consistency conditions 
which follow from matching to an appropriate supergravity theory, the low 
energy theory at long distances. It is motivated in part by ideas taken from 
\cite{polwit,horwit,gimpol} and by the more unified 
description of self-duality and background fields that appears in the recent 
papers \cite{witten}.

In an earlier paper with Chen \cite{ccn} we obtained the short distance 
potential between heavy sources in a gauge theory in some generic background 
of the bosonic string. The potential is extracted from a covariant string
path integral representation of the pair correlation function of Wilson 
loops. Our results apply both in the 
background with $d$$=$$26$ spacetime dimensions, or in the presence of a 
generic background for the Liouville theory with fewer matter fields, $c_m$$<$$d$. 
We find an attractive, and scale invariant, $1/r$ short distance interaction
between the heavy gauge theory sources.
We note that the bosonic string has a tachyon, which must either be stabilized
(see the recent discussions in \cite{sen}), or 
eliminated--- as is possible in the type I and type II string theories. 
The bosonic results are a useful warm-up for the computation of the annulus 
in stable backgrounds of the superstring. They 
also capture the correct qualitative features of the short distance potential 
in a background of the superstring with a tachyon instability. The calculation
in the bosonic string proceeds as follows \cite{ccn}.

We consider heavy sources in the gauge theory in relative collinear motion with 
$r^2$$=$$R^2$$+$$v^2 \tau^2$, $v$$<<$$1$, thus giving a simple realization of
coplanar loops while mimicking 
straight-line trajectories in the Euclideanized $X^0$, $X^1$ plane.
Here $ r$ is their relative position, and 
$\tau$ is the zero mode of the Euclideanized time coordinate,
$X^0$. The scattering plane is wrapped into a spacetime cylinder by 
periodically identifying the coordinate $X^0$. Then the closed world-lines of 
the heavy sources are loops singly wound about this cylinder. We identify 
these closed world lines with Wilson loops. Following the earlier works of 
Alvarez \cite{alvarez}, and of Cohen et al \cite{cohen},
we give a path integral prescription for the pair correlation function
of Wilson loops. The loops can be taken to lie in the world-volume of a 
higher dimensional Dpbrane. Taking the large loop length limit of the 
correlation function, $L_i$$\simeq$$L_f$$\simeq$$ T$$\to$$\infty$, 
with $R$ held fixed, we define an effective potential as follows:
\begin{equation}
< M(C_i) M(C_f) > = -i \lim_{T\to\infty} \int_{-T}^{+T}
    d\tau V_{\rm eff.}[r(\tau),u]
\quad .
\label{eq:potenti}
\end{equation}
The dominant contribution to the potential between the sources at
short distances is from the massless modes in the open string 
spectrum. Suppressing the tachyon, and restricting to the massless 
modes of the bosonic string, the potential can be expressed as a 
systematic double expansion in small velocities and short distances 
with the result \cite{ccn}:
\begin{equation}
V_{\rm bos.} (r,u) = - (d-2) {{1}\over{r}} + O(z^2,\upz,u^2)
\quad ,
\label{eq:statdzero}
\end{equation}
where $z$ is the dimensionless scaling variable, $z$$=$$\rmin^2/r^2$, 
and $\rmin^2$$=$$2 \pi \ap u$ is the minimum distance scale 
probed in the collinear scattering of the heavy point sources.
The $1/r$ static term receives velocity dependent corrections which are
parameterized by the dimensionless variables, $z^2$, $\upz$, and $u^2$. 
We show that the small velocity short distance approximation is self-consistent for 
distances in the regime, $2\pi\alpha^{\prime}u$$<<$$r^2$$<<$$2 \pi\alpha^{\prime}$, 
and velocities in the range, $u$$<<$$u_+$, where the upper bound, $u_+$,
on permissible velocities can be estimated as described in \cite{ccn}. 
Thus, String/M theory predicts velocity 
dependent corrections to the potential between two heavy sources
in relative slow motion in a gauge theory, the numerical coefficients of
which are given by a systematic expansion.

Evidence of a distance scale shorter than the string scale was originally
found in the nonrelativistic scattering of D0branes \cite{bachas,dkps},
which gives a linear potential in the bosonic string theory \cite{ccn}. 
The D0branes are assumed to have fixed spatial separation in a direction
$X^{d-1}$, and in relative motion with nonrelativistic velocity $v$ in an 
orthogonal direction $X^d$. The static linear potential between a pair of 
bosonic D0branes corresponds to a shift in the vacuum energy density
relative to the background with no Dbrane sources due to a constant 
background electromagnetic vector potential \cite{leigh,bachas} with 
vanishing electric field strength \cite{witten}: $A^{\mu}$$=$${\bar A}^{\mu}$,
$\mu$$=$$d$$-$$1$, $\partial_0 {\bar A}^{\mu}$$=$$0$. 
The systematics of the small velocity short distance double expansion, and 
the value $r_{\rm min.}^2$$=$$2\pi\alpha^{\prime}u$ for the minimum distance 
probed in the scattering of D0branes, are in precise agreement with our 
results for the short distance potential between heavy gauge theory sources.

In this paper, we extend these results to the generic stable background 
of the supersymmetric type I and type II theories. We give a path integral
prescription for the supersymmetric annulus amplitude, determining both
its phase and its normalization from first principles. Boundary 
reparameterization invariance is imposed following the analysis of the 
bosonic string amplitude given in \cite{ccn}. Our prediction for the short 
distance potential between heavy gauge theory sources in supersymmetric 
string theory is:
\begin{equation}
V_{\rm super.}(r,u) = - {{u^4}\over{r^{9}}}  
2^{4} \pi^{7/2} {\alpha^{\prime}}^{4}
\Gamma({{9}\over{2}}) + O(u^6)
\quad ,
\label{eq:super}
\end{equation}
which can be compared with the bosonic string result given in 
Eq.\ (\ref{eq:statdzero}). The systematics of the velocity dependent 
corrections are much simpler in the fermionic string. This is a consequence
of the BPS conditions or, equivalently, as in this example, of spacetime 
supersymmetry. 
The key ingredient which enables a prediction of the numerical coefficient
in the short distance potential is its relationship to the vacuum
energy computation in string theory: unlike in quantum field theories, the
one-loop cosmological constant in critical string theory can be unambiguously
normalized, an observation due to Polchinski \cite{poltorus}.

We begin in section II with a brief discussion of supersymmetric boundary 
conditions and spin structure. Spinor conventions, and a recapitulation of the
local symmetries of the world-sheet action for the fermionic string, 
are given in appendix A. Section III contains 
an evaluation of the supersymmetric annulus from first principles, beginning 
with the covariant path integral over world-sheets of specified spin structure 
and specified boundary condition on all of the world-sheet fields. The gauge fixing of 
the local world-sheet symmetries is carried out in section IIIA where we give 
a derivation of the supersymmetric annulus with boundaries on parallel and 
static Dbranes, up to undetermined phases. The precise global structure of 
the gauge orbit of the supersymmetry and super-Weyl transformations on the 
world-sheet is not known 
\cite{dhokercomp,polchinskibook}. We will take the point of view that global 
ambiguities in the superstring path integral can be eliminated by the imposition
of infrared consistency conditions that require matching to an appropriate 
supergravity theory describing the low energy physics. This prescription for
the phases of the fermionic path integrals summed in the vacuum amplitude is
given in section IIIB, using simple, and universal, infrared consistency 
conditions on the long distance physics. The result is an unambiguous determination 
of both the normalization and the phase of the supersymmetric annulus. 
Implementing boundary reparameterization invariance in the path
integral as in \cite{ccn}, we derive an expression for the 
supersymmetric pair correlation function of Wilson loops in section IIIC.
Finally, we extend our results in section IIID to generic boundary conditions
corresponding to Dbranes in relative motion, imposing appropriate infrared consistency
conditions as before. As a consistency check, we compute the short 
distance potential probed in the nonrelativistic small angle scattering of
Dpbranes, recovering the numerical coefficient previously obtained 
in \cite{polchinskibook}.

In section IV we adapt these results to the supersymmetric pair correlation 
function of Wilson loops corresponding to worldlines of heavy sources in 
relative slow motion. We show that the short distance potential between heavy 
sources in a supersymmetric gauge theory takes the form of a scale invariant 
$1/r$ fall-off contributed by the bosonic degrees of freedom, multiplicatively 
corrected by a convergent power series expansion in the dimensionless variable 
$z$$=$$r_{\rm min.}^2/r^2$. The leading term in the potential is given by the
expression in Eq.\ (\ref{eq:super}). Some implications of our results are 
sketched in the conclusions.

\section{Supersymmetric Boundary Conditions and Spin Structure}
\label{sec:boundary}

We begin with the Brink-Di Vecchia-Howe-Deser-Zumino world-sheet action 
\cite{dzbdh} used in Polyakov's path integral quantization of the fermionic 
string \cite{polyakov}:
\begin{equation}
S_{SP} = \halft \int d^2 \sigma \sqg [ \half g^{mn}\partial_m \Xmu \pn X_{\mu} 
+ \half \ap \psib^\mu \gm \partial_m \psi_\mu + \sqap (\chib_a \gm\ga \psi^{\mu})
  (\partial_m X_{\mu}) + \apfour (\chib_a\gb\ga\psi^{\mu})(\chib_b\psi_{\mu}) ]
\quad ,
\label{eq:local}
\end{equation}
invariant under both reparameterizations and local 
$N$$=$$1$ world-sheet supersymmetry transformations.
The indices $m$,$n$$=$$1$,$2$ label the world-sheet coordinates for the string 
with metric $g$, and 
$a$,$b$$=$$1$, $2$ label the flat local tangent space to the world-sheet. 
Spinor conventions and the local symmetries underlying the action are reviewed in 
the appendix. We use the label $\mu$$=$$0$, $\cdots$, $p$ for the Neumann directions 
parallel to the worldvolume of the Dpbranes. The branes are spatially separated 
by $R$ in the $X^9$ direction, with $\mu$$=$$p$$+$$1$, $\cdots$, $9$ labeling the 
Dirichlet directions orthogonal to the worldvolume of the Dpbranes. 
The boundary conditions on the embedding coordinates, $X$, and their fermionic
superpartners, $\psi$, are obtained from the kinetic term in 
Eq.\ (\ref{eq:local}). The corresponding free field action in a 
flat embedding spacetime with component fermions, $\psi^{\pm}_{\mu}$,
is \cite{polchinskibook}:
\begin{equation}
S[X,\psi] = \quaft \int d^2 \sigma [ \partial^m \Xmu \partial_m X_{\mu} + 
    \ap ( \psi^{-\mu} (\pone - i \ptwo) \psimi_\mu +
   \psi^{+\mu} (\pone + i \ptwo ) \psipl_\mu ) ] 
\quad ,
\label{eq:faction} 
\end{equation}
which extends to the locally supersymmetric action given above. A variation of the 
classical action with respect to the embedding coordinate $X$ gives the surface term:
\begin{equation}
\delx (n^a \partial_a X_{\mu} ) = 0 
\quad .
\label{eq:varbaction} 
\end{equation}
As possible boundary conditions we list: 
\begin{eqnarray}
{\rm N} \quad  &&: \quad n^a \pa \Xmu =0 
\nonumber \\
{\rm D} \quad &&: \quad \Xmu = y^{\mu} 
\nonumber \\
{\rm W} \quad  &&: \quad \delx \propto t^a \pa \Xmu  
\quad , 
\label{eq:bc}
\end{eqnarray}
where $y^{\mu}$, $\mu$$=$$p+$$1$, $\cdots$, $9$,
gives the spacetime location of the Dbrane.
The $W$, or modified Dirichlet (MD), boundary condition is motivated by the 
Wilson loop problem \cite{alvarez,ccn}. It permits fluctuations in the 
world-sheet fields tangential to the boundary. The 
boundaries of the world-sheet have been identified with the
closed world-lines of a heavy quark--antiquark pair in the gauge 
theory.

A point source undergoing straight line motion with nonrelativistic velocity $v$ 
in the $X^0$, $X^1$ plane with respect to the origin, $X^0$$=$$X^1$$=$$0$, 
and in zero external field, is described by the boundary conditions:
 \begin{equation}
{\rm V} \quad : \quad n^a \pa (X^{0} - v X^1 ) = 0 , \quad \quad X^1 = v X^0 \quad ,
\label{eq:gbc} 
\end{equation}
with N (D) boundary conditions imposed on the $X^0$ $(X^1)$ coordinates of the 
source fixed at the origin. The boundary conditions on two point
sources in relative 
motion in a ${\cal D}$-manifold 
are rather simple, irrespective of whether the motion occurs within the 
worldvolume of a higher dimensional Dbrane, the intersection of two or more
Dbranes, or in the bulk transverse space orthogonal to some configuration of Dbranes.
The point sources are the end-points of open strings.
Then we distinguish the 
$d$$=$$10$ embedding coordinates of the world-sheet as NN, ND, or DD, directions, 
depending on whether both, one, or neither, point source
has nonvanishing spacetime momentum in the direction of the coordinate 
\cite{dbrane,polchinskibook}. In the discussion that follows, 
we restrict ourselves to NN and DD coordinates alone. Note that identical
boundary conditions must be imposed on all of the NN, and all of the DD, 
fermions in order to preserve the global $SO(1,p)$$\times$$SO(9-p)$ symmetry 
of the vacuum amplitude. The world-sheet gravitino satisfies the same boundary 
conditions as the NN fermions--- this is dictated by supersymmetry.

Consider the variation of the world-sheet action with respect to the 
fermion field. The vanishing of the surface term dictates the following 
condition on fermion bilinears on the boundary:
\begin{equation}
\psi^{+\mu}(\delta \psi_{+\mu}) = \psi^{-\mu}(\delta \psi_{-\mu})  \quad , 
\label{eq:bf}
\end{equation}
with solutions, $\psi^{+\mu}$$=$$\pm \psi^{-\mu}$, at any boundary. As a
check, we perform a supersymmetry transformation on the surface term. 
This gives the condition, $ n_a[ (\pb X_{\mu})( \xib \ga\gb\psi^{\mu})] 
$$=$$ 0 $. Align the world-sheet with coordinate $\stwo$ normal, and 
$\sone$ tangential, to the boundary. Assume Neumann boundary conditions 
on the embedding coordinate $X_{\mu}$: 
$\ptwo X_{\mu}$$=$$0$, with $\pone X_{\mu}$$\neq$$0$.
Then the requirement that the boundary conditions on the $\psi^{\mu}$ 
preserve world-sheet supersymmetry implies the restriction, 
$\xi^+ $$=$$\mp \xi^-$, on permissible supersymmetry transformations 
on the boundary. With the Dirichlet boundary condition, $\pone X_\mu$$=$$0$, 
we obtain the constraint, $\xib \psi^{\mu}$$=$$0$. Thus, choosing one or other
sign for the NN fermions simultaneously determines the choice of phase for
the DD fermions. In component form,
the choice $\xi^+$$=$$\mp \xi^-$ implies that $\chi^{-\mu}$$=$$\mp \chi^{+\mu}$. 
The constraints on the fermionic fields for the W and V boundary conditions 
on the world-sheet are a straightforward generalization of this reasoning.

The annulus 
amplitude in a flat spacetime background of the supersymmetric string can be 
represented as a path integral summing over fluctuations of world-sheets with 
cylindrical topology weighted by the action in Eq.\ (\ref{eq:local}): 
\begin{equation}
\A = \half \sum_{\beta,\alpha \in 0,1} \C 
   \int_{[\beta,\alpha]} {{[dX][d\psi][d g][d\chi]}\over{{\rm Vol}(gauge)}} 
e^{-S_{SP}[X,\psi,g,\chi] - \mu_0 \int_{\cal M} \ds \sqg - S_{\rm ren.}} 
\quad .
\label{eq:path}
\end{equation}
Thus, the world-sheets of the fermionic string are endowed with additional degrees 
of freedom, and the quantum fluctuations about some minimum action configuration 
summed
in the path integral must include a consideration of these modes. We consider the 
simplest background configuration of static parallel Dpbranes separated by a 
distance
$R$. The stretched string between the Dbranes 
contributes a term, $-R^2l/4\pi\alpha^{\prime}$, 
to the action 
in Eq.\ (\ref{eq:path}), further corrections being suppressed at weak coupling. 
Since 
the one-loop vacuum amplitude is a sum over surfaces of 
cylindrical topology with Euler characteristic $\chi$$=$$0$, the amplitude is free of any
dependence on the string coupling constant. Also, we can drop boundary
cosmological constant terms in favor of the bulk cosmological constant, $\mu_0$, 
since these
are not independent Lagrange parameters on the cylinder. We will gauge all of the 
local symmetries reviewed in appendix A.
$S_{\rm ren.}$ contains any additional
counterterms that may be necessary in order to obtain amplitudes invariant under both
world-sheet diffeomorphisms and Weyl transformations of the metric, as well as local 
supersymmetry transformations and super-Weyl rescalings of the world-sheet gravitino. 
Divergent contributions to the path integral arising from local gauge anomalies in
the measure will be absorbed in a renormalization of the bare couplings introduced 
in $S_{\rm ren.}$, including the bulk cosmological constant, the renormalized
values being set to zero at the end of the calculation \cite{poltorus,cohen}.

We turn next to global aspects of the path integral. We are summing over
spin structures and averaging over the $\pm$ ambiguity in the boundary condition
on the fermions at the boundary of the world-sheet. The label $\alpha$ 
on the path integral refers to a choice of spin structure on the world-sheet: 
the change in phase in a Weyl fermion upon traversal of a closed path 
homotopic to either boundary of the cylinder. Under $\sigma^1$$\to$$\sigma^1+1$,
the left and right-moving component fermions transform as:   
\begin{equation}
\psi^{\pm \mu} (\sone + 1, \stwo ) = - \ssa \psi^{\pm \mu} (\sone , \stwo ) \quad . 
\label{eq:sigma1}
\end{equation}
The parameter
$\alpha$$=$$0$ ($1$) labels the string path integral computed with world-sheet 
spinors that are, respectively, anti-periodic (periodic) around the single 
closed cycle of the cylinder. The parameter $\beta$ 
denotes the ambiguity, $\psi^{+\mu}$$=$$\pm \psi^{-\mu}$, at any boundary,
described above. Specifically, for either NN or DD fermions, we will define 
$\beta$ as follows: 
\begin{equation}
\psi^{+ \mu} (\sone , 0) = - \ssb \psi^{-\mu} (\sone , 0) , \quad \quad 
{\rm with} ~ \psi^{+ \mu} (\sone , 1) = \psi^{-\mu} (\sone , 1) \quad . 
\label{eq:sigma2}
\end{equation}
Choosing the phase at the $\stwo$$=$$1$ end-point to correspond to periodic 
fermions is pure convention. Thus, $\beta$$=$$0$ $(1)$
corresponds to reflection with (without)
a phase change of $\pi$ in the fermionic wavefunction
at the $\sigma^2$$=$$0$ end-point.
We remark that this convention corresponds to that in the text 
\cite{polchinskibook}. In the critical dimension, and with an unambiguous 
prescription for 
the phases, $\C$, of the path integrals, there will be no global gravitational or 
Lorentz anomalies.
Our prescription for determining the
absence of global phase ambiguities in the one-loop vacuum amplitude will be 
physical, rather than 
constructive. It is motivated by infrared consistency conditions on the long 
distance physics, as will be clarified in section IIIB. For the 
discussion in section IIIA, we encourage the reader to think of 
the $\C$ as {\em unspecified}, and therefore
ambiguous, phases that weight the different contributions to the annulus amplitude.

\section{Path Integral Evaluation of Supersymmetric Annulus}
\label{sec:annulus}

We will now give a path integral evaluation of the supersymmetric annulus---
a sum over orientable world-sheets with boundaries on parallel Dpbranes, 
paying special attention to the imposition of world-sheet supersymmetry 
both in the bulk, and on the boundary. We begin with the simplest configuration of
parallel and static Dpbranes, generalizing our results in section IIID for the 
$V$ boundary conditions describing Dpbranes in relative
motion. The gauge fixing of the local symmetries on the world-sheet, and
a derivation from first principles of the annulus amplitude up to unknown phases, is
described in section IIIA. 
In section IIIB, we show how infrared consistency conditions can be used to 
determine all of the phase ambiguities in the path integral.
We extend this derivation in IIIC to an analysis of the supersymmetric pair
correlation function of Wilson loops following the treatment in \cite{ccn}. 
The extension to generic boundary conditions describing branes in relative motion
is given in section IIID. This result will be adapted in section IV to give an
expression for the supersymmetric pair correlation function of Wilson loops 
corresponding to the worldlines of heavy gauge theory sources in slow relative 
motion. The normalization and the phase of the pair correlation function is 
therefore precisely, and unambiguously, determined, allowing a derivation of
the short distance potential between the sources.

\subsection{Gauge Fixing of the Local World-sheet Symmetries}
\label{sec:gauge}

We begin by gauge fixing the Lorentz transformations in the local tangent space at
any point in the world-sheet, thereby eliminating one of the four bosonic gauge 
parameters. This implies that, although it is convenient to write classically 
covariant expressions in terms of zweibeins $\ema$, the number of physical degrees 
of freedom in the path integral is the same as with metrics: we use local Lorentz 
rotations to eliminate one of the independent degrees of freedom in the zweibein. 
Likewise, we eliminate two of four independent modes of the gravitino by choosing 
super-conformal gauge, $\gamma^m\chi_m$$=$$0$ \cite{howe}, thereby gauge fixing 
super-Weyl transformations. 
This immediately
creates an apparent problem with supersymmetry since we have only two fermionic
but three bosonic gauge parameters, but this is not so. Recall that we work in the
critical spacetime dimension gauging Weyl transformations of the metric. As in the
path integral quantization of the bosonic string \cite{poltorus},
although it is convenient to keep the Weyl mode in the discussion of the measure,
the principle of ultralocality requires that any explicit dependence on the Weyl mode
only contribute local, renormalizable, terms to the effective action
\cite{poltorus,cohen}. The unique choice for such terms is the Liouville action.
Thus, in the critical dimension, the Weyl mode entirely decouples.

We will employ similar reasoning in gauging local Lorentz and super-Weyl 
transformations: ultralocality of the measure in the path integral
\cite{poltorus} requires that
any explicit dependence on the Weyl and super-Weyl modes can only
contribute terms proportional 
to the supersymmetric Liouville action \cite{polyakov}.
Thus, although we find it 
convenient to keep all four bosonic and fermionic gauge 
parameters in the discussion below, there are 
really half as many physical gauge parameters in the critical superstring corresponding, 
respectively, to diffeomorphisms and local supersymmetry transformations. Following 
gauge fixing, the path integrals reduce to ordinary integrals over constant modes and
moduli. The counting of superconformal Killing spinors and supermoduli on a 
Riemann surface is given by the supersymmetric analog 
of the Riemann-Roch theorem. For cylindrical topology, the answer is rather simple,
since both the superconformal Killing spinors and the supermoduli are simply constant
spinors. These can only exist on a cylinder with both periodic spin structure and 
the periodic boundary condition at the endpoints of the open string.

The zweibein and metric are related by $g_{mn}$$=$$\ema\enb \delta_{ab}$. We make 
the same choice of fiducial world-sheet metric as in the analysis of the bosonic 
string:
\begin{equation}
ds^2=l^2 (d\sone)^2 + (d\stwo)^2 , \quad \quad 0 \le \sone \le 1, \quad 0 \le \stwo \le 1
\quad .
\label{eq:cylmet}
\end{equation} 
Thus, the fiducial einbein on the boundary is ${\hat e}$$=$$\sqg$, and the 
modulus $l$ can be identified as the boundary length, $l$$=$$\int_0^1 d \sone {\hat e} $ 
\cite{poltorus,cohen,ccn}. With this choice, the area of the fiducial 
world-sheet, $\area$, equals $l$.

Begin with the integration over the bosonic embedding coordinates, $\Xmu$. 
Integration over $p$$+$$1$ constant modes in the Neumann directions, and 
upon normalizing the measure for infinitesimal variations as in \cite{poltorus} 
we obtain the result: 
\begin{equation}
   \int [dX] e^{-S_{SP}[X,\psi,g,\chi] }
= 2 iV_{p+1} (4\pi^2 \alpha^{\prime})^{-\ponehalf} l^{\ponehalf}
e^{-R^2 l/4\pi\alpha^{\prime}} 
 ({\rm det}^{\prime} \Delta )_g^{-5}  e^{-S_{\rm eff.}[\psi,\chi,{\hat g}]} \quad, 
\label{eq:xint}
\end{equation}
where the determinant for the scalar Laplacian is computed with the NN boundary
condition on $p$$+$$1$, and DD boundary condition on $9$$-$$p$, coordinates. The 
overall factor of 
two accounts for the two possible orientations of the open string. The 
effective action for the fermionic fields takes the form:
\begin{eqnarray}
S_{\rm eff.} =&&  \halft \int d^2 \sigma \sqg 
[ \half \ap \psib^\mu \gm 
\partial_m \psi_\mu + \aptwo (\chib_a\gb\ga\psi^{\mu})(\chib_b\psi_{\mu}) ]
\nonumber\\
&& \quad - \int d^2 \sigma \sqg (\chib_a \gm\ga \psi^{\mu})(\sigma) 
\int d^2 \sigma^{\prime} \sqg (\chib_a \gamma_m \ga \psi_{\mu}) (\sigma^{\prime}) 
\partial^2_{\sigma\sigma^{\prime}} G(\sigma,\sigma^{\prime})
\quad .
\label{eq:eff}
\end{eqnarray}
$G(\sigma,\sigma^{\prime})$ is the Greens function of the scalar Laplacian.

Consider now the gauge fixing of the local symmetries of the effective action. 
A bosonic deformation of the metric is decomposed into a Weyl transformation,
a diffeomorphism continuously connected to the identity, and a change in the
length of the boundary of the cylinder. We will gauge both diffeomorphisms and
Weyl transformations of the metric. Defining implicitly the measure for
infinitesimal variations in the tangent space to the space of metrics as in
\cite{poltorus} gives the result:
\begin{equation}
{{1}\over{{\rm Order}({\tilde{D}})}}
\int {{[dg] ({\rm det}^{\prime} \Delta )_g^{-5} }
  \over{ {\rm Vol}({\rm Diff}_0 \times {\rm Weyl}) }} 
= {{1}\over{2}} \int {{[d\delv]}\over{{\rm Vol}({\rm Diff})_0}}
\int {{[d\delta \phi] }\over{ {\rm Vol}({\rm Weyl}) }}
 \int_0^{\infty} dl ~ J_b(l;{\hat g}) e^{-{{10-d}\over{48\pi}} S_L[\phi,g]} 
   ({\rm det}^{\prime} \Delta )_{\hat g}^{-5} 
  \quad ,
\label{eq:Jacobm} 
\end{equation}
where $J_b$ is the Jacobian from the change of variables 
computed in \cite{poltorus}:
\begin{equation}
J_b={{(l/2\pi)^{1/2} \cdot ({{2}\over{l^2}})^{1/2} \cdot
(\half l^2 \eta^4(\lh ))^{1/2} }\over{(l^3/2\pi)^{1/2}}} \quad ,
\label{eq:jacobb}
\end{equation}
in the cylinder metric specified above. We have divided by the order of the subgroup
of the disconnected component of the diffeomorphism group, ${\tilde {D}}$:
discrete diffeomorphisms of the world-sheet left invariant under the choice
of superconformal gauge \cite{poltorus}. This gives a factor of two in the 
denominator of Eq.\ (\ref{eq:Jacobm}),
correcting for the two-fold invariance of the measure under the 
diffeomorphism: $\sigma^1$$\to$$-\sigma^1$. The Weyl anomaly of the measure 
exponentiates to a term proportional to the Liouville 
action, whose coefficient vanishes in the critical spacetime dimension, $d$.
Consider the result of performing a local supersymmetry and super-Weyl
transformation on this expression. This will induce a super-Weyl anomaly. We must
simultaneously include the contributions from
world-sheet fermions to the Weyl anomaly.

An arbitrary fermionic deformation of the world-sheet gravitino can be decomposed:
\begin{equation}
\delchi = - D_m \delxi + (\delzeta)\gamma_m  + (\partial_{\alpha}
\chi_m) (\delta \nu^{\alpha} ) \quad ,
\label{eq:gravitino}
\end{equation} 
where $\delxi$ is an infinitesimal supersymmetry transformation, $\delzeta$ 
is a rescaling of the gravitino, and $\delta\nu$ is a change in a possible 
supermodulus--- a constant two component spinor, $\nu$, on the cylinder.
We work in superconformal gauge, invoking super-Weyl transformations in setting 
$\gamma^m\chi_m$$=$$0$ \cite{howe}. We make an orthogonal decomposition into 
infinitesimal deformations parallel (perpendicular) to the gauge slice, 
respectively, preserving (violating) the restriction to gamma traceless 
$\chi_m$. In addition, we must separate the contribution from superconformal
Killing spinors, $\delxi_0$, which leave the world-sheet gravitino unchanged:
\begin{equation}
-D_m (\delxi_0) + (\delzeta_0)\gamma_m = (- D_m + \half \gamma_m \gamma^n D_n)(\delxi_0) = 0  
\quad , 
\label{eq:scks}
\end{equation}
since the supermodulus, $\delzeta_0$$=$$\half \gm D_m (\delxi_0)$, is in 
the kernel of the operator $D_m$. The superconformal Killing spinor, 
$\delxi_0$, is in the kernel of the operator, 
$P_{1/2}$$=$$-2D_m $$+$$ \gamma_m \gn D_n$.
In the fiducial cylinder metric, $D_m$$=$$\partial_m$, and the Killing
spinor is simply a constant spinor. Likewise, the supermodulus is also
a constant spinor. Constant spinors can only exist on cylindrical 
world-sheets with both periodic spin structure {\em and} periodic boundary 
condition at the endpoints of the open string. Thus, for 
$\alpha$$=$$\beta$$=$$1$, we have one superconformal constant spinor and
one supermodulus to be accounted for in the measure. 
Performing the change of variables in Eq.\ (\ref{eq:gravitino}) 
we can write: 
\begin{equation}
\int_{[\beta,\alpha]}
 {{[d\delchi]}\over{{\rm Vol}({\rm sWeyl}\times{\rm sDiff})}} = 
  \int {{[d\delxi]^{\prime}}\over{{\rm Vol}({\rm sDiff})_0}} 
\int {{[d\delta \zeta] }\over{ {\rm Vol}({\rm sWeyl}) }} 
   \int_{[\beta,\alpha]} [d\delta\nu ] J_f^{(\beta,\alpha)}  
\quad ,
\label{eq:gravmes}
\end{equation}
where the integration over a supermodulus is absent unless 
$\alpha$$=$$\beta$$=$$1$.
A norm in the tangent space to the
space of the spin $3/2$ field $\chi_m$
invariant under both reparameterizations and local Lorentz
transformations
can be written as follows:
\begin{equation}
|\delchi|^2 = -i \int \ds \sqg (\delchinob^m) (\delchino_m) =
2 \int \ds \sqg (\delchino_1^+ \delchino_1^- + \delchino_2^+ \delchino_2^- ) 
\quad ,
\label{eq:norm} 
\end{equation}
where the measure in the path integral is normalized as in \cite{poltorus}:
\begin{equation}
1 = \int [d\delchi] e^{-|\delchi|^2/2} \equiv 
 \int [(d\delchino_1^+)(d\delchino_1^-)(d\delchino_2^+)(d\delchino_2^-)] 
 e^{-|\delchi|^2/2} \quad .
\label{eq:normg} 
\end{equation}
We can likewise define a norm in the tangent space to the space of 
$\xi$:
\begin{equation}
|\delxi|^2 = - i \int \ds \sqg (\delxib)(\delxi)  , \quad \quad 
1 \equiv \int [(d\delxi^+)(d\delxi^-)] e^{ - \int \ds \sqg (\delxi^+)(\delxi^- )} 
\quad .
\label{eq:normxi} 
\end{equation}
Separating the ordinary Grassmann integral over the superconformal 
Killing spinor, $\xi_0$, gives: 
\begin{equation}
1 = \int 
[d\delta \xi_0] e^{-\itwo (\delta{\bar \xi_0}) (\int \ds \sqg) (\delta{ \xi_0}) }
\int [(d\delxi^+)^{\prime}(d\delxi^-)^{\prime}] e^{-|\delxi|^2/2} 
= l \int [(d\delxi^+)^{\prime}(d\delxi^-)^{\prime}] e^{-|\delxi|^2/2} 
\quad .
\label{eq:normxisck} 
\end{equation}
Likewise, the ordinary Grassmann integration over a supermodulus gives:
\begin{equation}
\int [d\delta \nu ] e^{-\itwo (\delta{\bar \nu 
})( \int \ds \sqg )(\delta{\nu }) } 
= l \quad .
\label{eq:normxism} 
\end{equation}
This last term is only necessary when computing the path integral with 
$\alpha$$=$$\beta$$=$$1$.  The measure for the path integral with constant spinors 
on the worldsheet is discussed in appendix B. For a configuration of
parallel and static Dbranes, the presence of the constant mode on the
world-sheet gives a vanishing result for the path integral. Henceforth, we 
restrict our discussion to the cases $(\beta,\alpha)$$\neq$$(1,1)$, leaving a
discussion of the $\alpha$$=$$\beta$$=$$1$ results to appendix 
B. Note that the norms in Eqs.\ (\ref{eq:norm})--(\ref{eq:normxisck}), are neither 
Weyl, nor super-Weyl, invariant. Weyl and super-Weyl transformations generate
variations orthogonal to the gauge slice defined by gamma traceless $\chi$. 
Consequently, both symmetries are anomalous, giving contributions to the path 
integral measure which will exponentiate to terms proportional to the 
supersymmetric Liouville action \cite{polyakov}.

The result of a local supersymmetry 
transformation plus super-Weyl scaling on the expression in Eq.\ (\ref{eq:Jacobm}) is 
a variation outside the superconformal gauge slice: $S_L$ supersymmetrizes to the 
super-Liouville action and 
the super-Weyl anomaly exponentiates to a local renormalizable term proportional to 
the induced $\gamma^m(\delta\chi_m)$. The 
integration over diffeomorphisms continuously connected to the identity gives the 
volume of the Weyl group, cancelled by the term in the denominator. 
Thus, under a local supersymmetry and super-Weyl transformation, the 
variation of Eq.\ (\ref{eq:Jacobm}) gives:
\begin{equation}
\int {{[d\delta \phi] }\over{ {\rm Vol}({\rm Weyl}) }}
 \int_0^{\infty}  dl ~ J_b(l;{\hat{g}}) e^{-{{10-d}\over{48\pi}} S_{SL}[\phi,\zeta,g]} 
   ({\rm det}^{\prime} \Delta )_{{\hat{g}}}^{-5} 
  \quad ,
\label{eq:Jacobms}
\end{equation}
where $S_{SL}[\phi,\zeta,g]$ is the super-Liouville action. In the fiducial cylinder metric, 
\begin{equation}
S_{SL} [ \phi,\zeta, g ] = \int \ds \fidg [ 
\half \gmn \partial_m \phi \partial_n \phi + \zetab \gamma^m \partial_m \zeta 
+ \mu_1 e^{\phi} + \mu_2 \zetab \zeta e^{\phi/2} ] \quad , 
\label{eq:supLi}  
\end{equation}
where $\mu_1$ and $\mu_2$ are induced violations of, respectively, the Weyl
and super-Weyl invariance of the measure.

Likewise, consider the measure for the fermionic fields in the path integral. 
In the absence of a supermodulus, the effective action for the fermi fields 
given in Eq.\ (\ref{eq:eff}) reduces to the free field action for the matter 
fermions alone: the world-sheet gravitino can be entirely gauged away. As 
with the gauge fermions, there are no constant modes for the matter fermion
on a world-sheet of cylindrical topology when $\alpha$,$\beta$$\neq$$1$. 
Integrating over the $\psi^{\mu}$ with norm,
\begin{equation}
|\delpsi^{\mu}|^2 = -i \int \ds \sqg (\delpsib^{\mu})(\delpsi_{\mu})  , 
  \quad \quad 
1 = \int [(d\delpsi^{\mu+})(d\delpsi^{\mu -})] e^{- \int \ds \sqg 
   (\delpsi^{\mu+})(\delpsi^-_{\mu} )} 
\quad ,
\label{eq:normpsi} 
\end{equation}
gives $({\rm det}\gamma^m\partial_m)_{g}^{d}$, in the fiducial 
cylinder metric. Thus, for $(\beta,\alpha)\neq(1,1)$, we obtain:
\begin{equation}
\int {{ [d\chi_m] ({\rm det} \gamma^m \partial_m )_g^{10} }
  \over{ {\rm Vol}({\rm sDiff} \times {\rm sWeyl}) }} 
= \int {{[d\delxi]}\over{{\rm Vol}({\rm sDiff})}}
\int {{[d\delta \zeta] }\over{ {\rm Vol}({\rm sWeyl}) }} 
J_f(l;{\hat g}) 
   [{\rm det} (\gamma^m \partial_m )]_{\hat g}^{10} 
e^{- {{10-d}\over{96\pi}} S_{SL} [\phi,\zeta,g] } \quad . 
\label{eq:Jacobf}
\end{equation}
Combining the expressions in Eqs.\ (\ref{eq:supLi}) and (\ref{eq:Jacobf})
we see that, in the critical spacetime dimension, the Liouville, and super-Liouville, 
modes entirely decouple. The Jacobian $J_f$ describes the change of variables 
from $\delchi$ to $(\delxi,\delzeta )$ and was first computed in 
\cite{dhokercomp}. In the absence of a supermodulus and superconformal Killing 
spinor on the world-sheet, the fermionic Jacobian is simply:  
\begin{equation}
J_f =  ({\rm det} P_{1/2}^{\dagger} P_{1/2} )^{-1/2}  \quad , 
\label{eq:jacfgen} 
\end{equation}
where $P^{\dagger}_{1/2}$$=$$\partial_m$ on the cylinder, is the adjoint 
of the operator $P_{1/2}$. The extension for $\alpha$$=$$\beta$$=$$1$ is discussed in
appendix B. Combining with results from Eqs.\ (\ref{eq:xint}), (\ref{eq:jacobb}), 
we obtain the following expression for the amplitude:
\begin{equation}
\A = {{i}\over{2}} V_{p+1} (4 \pi^2 \alpha^{\prime})^{-\ponehalf} 
  \int_0^{\infty} {{dl}\over{l}} l^{-\ponehalf} 
e^{-R^2 l/4\pi\alpha^{\prime} } \eta(\lh )^{-8}  
\sum_{(\beta,\alpha) \neq (1,1)} \C 
[{\rm det} (\gamma^m \partial_m )]_{(\beta,\alpha)}^{8}
  \quad ,
\label{eq:pathamp} 
\end{equation}
where the contribution from two left-moving and two right-moving component 
fermions--- one each of which is a timelike fermion \cite{polchinskibook}, 
has been cancelled against the fermionic Jacobian, $J_f$. We are assuming
{\em identical} boundary condition, $\beta$, for all NN, and DD, fermions on 
a world-sheet with fixed spin structure $\alpha$.

The fermionic determinants in Eq.\ (\ref{eq:pathamp}) can be computed using
the method of zeta function regularization. The component world-sheet
fermions are complexified into Weyl fermions, equivalently, using
bosonization, into chiral bosons \cite{polchinskibook}. 
An advantage is that the result 
can be readily generalized to the modified boundary conditions describing 
a pair of Dpbranes rotated relative to each other. Upon analytic continuation 
of a Euclidean embedding coordinate to Minkowskian time, this is equivalent
to imposing boundary conditions
describing parallel Dpbranes in relative motion. We begin by grouping the eight 
left-moving component fermions, $\psi^{+i}$, $i$$=$$1$, $\cdots$, $8$, into four 
left-moving Weyl fermions:
\begin{equation}
\psi^{+1}+i\psi^{+2}, ~  
\psi^{+3}+i\psi^{+4}, ~  
\psi^{+5}+i\psi^{+6}, ~  
\psi^{+7}+i\psi^{+8} \quad .   
\label{eq:complex}
\end{equation}
Likewise, complexifying eight right-moving component fermions, $\psi^{-i}$,
gives four right-moving Weyl fermions. This is the world-sheet fermion 
content of the fermionic string with both left and right moving $N$$=$$1$
world sheet supersymmetries. The open string boundary condition in Eq.\
(\ref{eq:sigma2}) reduces the number of independent fermionic degrees of 
freedom by half, since it relates corresponding left and right moving Weyl
fermions at the end-points. Thus, while the $4$$+$$4$ Weyl fermions are not 
all independent world-sheet fields, this is a convenient basis in which to 
express the result. We will work with the type I theory in its T-dual 
formulation with a type IA, or type IB, supersymmetry, as determined by 
Dpbranes with even, or odd, $p$$\le$$9$. For convenience, we will keep the full 
$SO(8)$ global symmetry of the transverse coordinates, imposing identical boundary 
conditions and spin structure on all four independent Weyl fermions on the 
world-sheet. The 
required chiral determinants can all be obtained from the single functional 
determinant:
\begin{equation}
{\rm det}^{\prime} \Delta^{(\beta,\alpha)}  = \prod_{n_1,n_2}^{\prime}
   ({{4\pi^2}\over{l^2}}) \left [ (n_1+ (\alpha - 1)/2 )^2 + 
     {{l^2}\over{4}}(n_2+ (\beta - 1)/2 )^2 ) \right ] \quad .
\label{eq:determ}
\end{equation}
computed by the method given in \cite{poltorus,cohen,ccn}. The chiral 
determinants are formally defined by taking a square root. The 
result is therefore ambiguous up to a phase. We retain the phase ambiguity, 
absorbing it in the unknown $\C$. For anti-periodic (periodic) 
Weyl fermions with $\beta$$=$$0 (1)$, and world-sheets with spin structure, 
$\alpha$, we obtain the result \cite{poltorus,polchinskibook,ccn}: 
\begin{equation} 
{\rm det}^{\prime} \Delta^{(\beta,\alpha)}  = 
 \left [ q^{\evac} \prod_{m=1}^{\infty} (1+ e^{\pi i \alpha} 
    q^{m - \mmi})(1+ e^{-\pi i \alpha} q^{m - \mpl}) 
  \right ] \equiv {{ \Theta _{(\beta,\alpha)} (0 , \lh )} 
   \over{\eta(\lh ) }} \quad .
\label{eq:alpha}
\end{equation}
As is shown in appendix B, in the case $\alpha$$=$$\beta$$=$$1$, the left hand 
side of Eq.\ (\ref{eq:alpha}) is corrected by a contribution from the measure 
which vanishes for boundaries on static Dbranes. Inserting the expressions 
for $(\beta,\alpha)$$\neq$$(1,1)$ in Eq.\ (\ref{eq:pathamp}) gives  
\cite{dhokercomp,polchinskibook}:
\begin{equation}
\A = {{i}\over{2}} V_{p+1} (8 \pi^2 \alpha^{\prime})^{-\ponehalf} 
\int_0^{\infty} {{dl}\over{l}} e^{-R^2 l/2\pi\alpha^{\prime}}
l^{-\ponehalf} \eta(il)^{-12}  
\left [ C^{0}_{0} \Theta _{(0,0)}^4 (0 , il ) 
+ C^{0}_{1} \Theta _{(0,1)}^4 (0 , il ) 
+ C^{1}_{0} \Theta _{(1,0)}^4 (0 , il ) \right ] 
  \quad ,
\label{eq:resamp} 
\end{equation}
where the $\C$ are undetermined phases. Note that we have 
rescaled by a factor $l$$\to$$2l$ in writing Eq.\ (\ref{eq:resamp}).

\subsection{Infrared Consistency of the Supersymmetric Annulus} 
\label{sec:phases}

We now come to the interesting issue of determining the phase of the path integral. 
The discussion that follows is based on ideas taken from \cite{polwit} and also 
\cite{horwit,gimpol,polchinskibook}. We will show that the following infrared 
consistency conditions:

\begin{itemize}
\item the elimination of the tachyon 

\item the absence of a static force between the Dbranes 

\end{itemize}

\noindent determine two of the three unknown phases in the expression for
the supersymmetric annulus given in Eq.\ (\ref{eq:resamp}). 
It should be emphasized at the outset that these 
requirements are {\em insufficient} to ensure infrared finiteness of the 
perturbative string theory \cite{polwit,polchinskibook}. The reason is 
that the oriented open and closed supersymmetric string has tadpoles for 
both the dilaton and the Ramond-Ramond scalar fields which are cancelled 
by contributions to the vacuum amplitude from non-orientable world-sheets 
in the full unoriented type ${\rm I}^{\prime}$ string theory 
\cite{gs,polcai,polchinskibook}. More generally, the presence of classical 
sources in the generic String/M theory background akin to the orientifold 
planes of the unoriented string may provide a means to cancel the 
troublesome tadpoles so we will leave this option open. Tadpole cancellation is 
an essential requirement for an infrared finite theory 
\cite{polwit,gimpol}. This is sometimes phrased as the requirement 
of BPS charge conservation in a compact space. 
We note that we are taking the point of view that an acceptable 
theory should have a sensible definition both in a noncompact, 
and a compactified, spacetime. In the compact 
space, the flux lines of the associated background field are required to 
close on a configuration of classical Ramond-Ramond sources. 
While we believe it likely that the infrared 
consistency conditions on the supersymmetric annulus amplitude 
listed above are necessary conditions which must be met
by the generic stable background of String/M theory--- irrespective 
of whether this is a background described by orientable or 
non-orientable world-sheets, we should note here the recent work on 
unstable brane configurations \cite{senhar},
and on the stabilization of the tachyon in open string
field theory \cite{sen}. Note also that we have distinguished the 
BPS condition---the absence of a static force between BPS sources 
in this example, from the specification of the spacetime 
supersymmetry. Each of these criteria impacts distinct 
renormalizability properties of the string theory \cite{polchinskibook}.

The third phase in the annulus amplitude, which we 
choose to be $C^0_0$, will be determined by computing the 
static limit of the long distance potential 
between Dbranes. Define the Minkowskian potential: 
\begin{equation}
\A(r,u) = -i \lim_{T\to\infty} \int_{-T}^{+T}
    d\tau V [r(\tau),u]
\quad ,
\label{eq:potentil}
\end{equation}
where $r^2$$=$$R^2$$+u^2 \tau^2$, and take the static limit $u$$=$$0$ 
\cite{bachas,dkps,polchinskibook,ccn}. This is 
a special case of the calculation that follows in section IIID.
Due to the no-force condition for the BPS configuration of static
and parallel Dbranes the amplitude vanishes. However, if we isolate the
contribution from the $(\beta,\alpha)$$=$$(0,0)$ sector alone, we can extract the
static long-range Newtonian gravitational potential between the Dbranes.
This is required to be attractive, which determines the phase $C_0^0$.
Thus, simple, and universal, infrared consistency conditions 
determine all of the unknown phases in the supersymmetric annulus.

Let us carry out this procedure explicitly. The long distance physics 
of the vacuum amplitude is dominated by the lowest lying closed 
string modes. This can be exposed more clearly 
by taking the $l$$\to$$0$ limit of the expression in Eq.\ 
(\ref{eq:resamp}). Using standard identities for the Jacobi theta functions 
\cite{ww,polchinskibook} gives:
\begin{eqnarray}
\A (r)
 =&& {{i}\over{2}}
 V_{p+1} \int_0^{\infty} {{dl}\over{l}} (8 \pi^2 \alpha^{\prime} )^{-\ponehalf} 
e^{-r^2 l/2\pi\alpha^{\prime} } l^{(7-p)/2} \eta(\li )^{-12} 
 \sum_{(\beta,\alpha)\neq(1,1)} ~  \C ~ 
   \Theta _{(\alpha,\beta)}^4 (0 , \li ) 
\nonumber 
\\ 
 =&& {{i}\over{2}} V_{p+1} (8 \pi^2 \alpha^{\prime} )^{-\ponehalf} 
\int_0^{\infty} dl e^{-r^2 l/2\pi\alpha^{\prime} } 
  l^{(5-p)/2} q^{-1/2} 
\nonumber 
\\ 
&& \quad \times \left \{ q^0 (C_0^0 + C_0^1 ) + \qh [8 
(C_0^0 - C_0^1 ) + 16 C_1^0 ] + O(q) \right \} 
\quad ,
\label{eq:vacuinv}
\end{eqnarray}
where $q$$=$$e^{-2 \pi/l}$.
The absence of the closed string tachyon appearing at $O(q^{-1/2})$ requires 
$C_0^1$$=$$-C_0^0$. The absence of a static force between the 
Dbranes requires that we set $C_1^0$$=$$-C_0^0$. The 8+8 massless states 
in the $(0,0)$, 
$(0,1)$ sector contribute to the vacuum amplitude with the opposite spacetime 
statistics of the 16 massless states in the $(1,0)$ sector. Thus, we discover 
an underlying spacetime supersymmetry in the vacuum amplitude 
although we have not required it. While the 
absence of the static force and the requirement of spacetime 
supersymmetry are equivalent conditions in this example--- the T-dualized
type I string in a background of parallel and static Dpbranes, the 
distinction may be of consequence for non-BPS brane configurations.

It remains to determine the overall phase of the annulus.
The massless states in the Neveu-Schwarz sector with spin 
structure, $(\beta,\alpha)$$=$$(0,0)$, contribute a static 
long range Newtonian interaction between the Dbranes, analogous to that 
in the bosonic string \cite{bachas,dkps,ccn}. It should be emphasized that
this term is universally present in the one-loop vacuum amplitude
of {\em any} fermionic string theory irrespective of background,
prior to possible cancellation by additional contributions to the 
amplitude. The sign of the potential is determined by the 
phase $C^0_0$. Defining the Minkowskian potential 
as in Eq.\ (\ref{eq:potentil}) and substituting 
for the phases in Eq.\ (\ref{eq:vacuinv}) gives:
\begin{eqnarray}
V^{(0,0)}_{\rm static} (r) 
   &&= - C_0^0 ~ 2 V_{p} (8 \pi^2 \alpha^{\prime} )^{-\ponehalf} 
 \half \int_0^{\infty} dl e^{-r^2 l/2\pi\alpha^{\prime} } 
  l^{(5-p)/2} 
     [ 8 + O(q^{1/2}) ]  
\nonumber
\\
 &&= - 8 C_0^0 ~  V_{p} (8 \pi^2 \alpha^{\prime} )^{-\ponehalf} 
           {{1}\over{r^{7-p}}} \Gamma ({{7-p}\over{2}}) 
              (2\pi \alpha^{\prime})^{(7-p)/2} 
+ \cdots 
\quad . 
\label{eq:polong}
\end{eqnarray} 
The factor of eight counts the transverse polarizations of the 
${\bf 8_v}$ multiplet under the $SO(9,1)$ Lorentz group. The potential
has a static remnant upon setting $u$$=$$0$. The Newtonian 
potential is required to be {\em attractive}, and we can 
therefore set $C_0^0$$=$$1$. Thus, 
\begin{equation} 
V_{\rm static}^{(0,0)} (R)= 
- (d-2) \cdot {{1}\over{R^{7-p}}} V_{p} 2^{2-2p} \pi^{(5-3p)/2} 
{\alpha^{\prime}}^{3-p} \Gamma ({{7-p}\over{2}}) \quad ,  
\label{eq:potl} 
\end{equation} 
where $R$ is the static separation of the Dpbranes.
We emphasize once again that the static interaction in 
Eq.\ (\ref{eq:potl}) will be {\em cancelled} by contributions to the 
vacuum amplitude from states in the Ramond sector and, in the full 
string theory, from the unoriented world-sheets.
Nevertheless, it has a simple physical interpretation which allows us
to use it to determine an unknown phase in the string path integral.

We have shown that infrared consistency conditions determine
all three phases in Eq.\ (\ref{eq:resamp}),
$C_0^0$$=$$1$, $C_1^0$$=$$C_0^1$$=$$-1$, giving
an unambiguous expression for the supersymmetric annulus. 
We emphasize that the ambiguity in phase has been determined by requiring
consistency with known {\em qualitative} features of the long distance 
physics. On the other hand, the normalization of the string path integral 
is unambiguously determined \cite{poltorus} leading to the prediction of
the numerical coefficient in Eq.\ (\ref{eq:potl}). We will use these 
expressions in section IV to make predictions about the short distance 
physics.

\subsection{Pair Correlation Function of Wilson Loops}
\label{sec:wilson}

In \cite{ccn}, we gave a path integral prescription for the pair correlation
function of macroscopic loop observables, $M(C_i)$, $M(C_f)$ in the weakly
coupled bosonic string theory, following the earlier work of Cohen et al 
\cite{cohen}. The loops, $C_i$, $C_f$, are taken to lie in a flat
D-manifold, ${\cal D}$, and are identified with the closed world-lines of 
heavy point sources in the gauge theory. The result for the short distance 
potential is independent of whether ${\cal D}$ is the worldvolume of 
some higher dimensional Dpbrane, the intersection of Dpbranes, or the 
bulk transverse space orthogonal to some configuration of Dbranes. The 
key issue is the implementation of boundary 
reparameterization invariance in the covariant one-loop string path integral.
Our interest is in the large loop length limit where the dynamics should be
universal, independent of the detailed geometrical parameters of the loops. 
In \cite{ccn}, we point out that the large loop length dynamics for generic 
loops is captured rather simply by summing over reparameterizations of loops 
with one or more marked points. For such maps the sum over 
reparameterizations 
of the boundary, $\partial M$, can be easily implemented in closed form even
{\em prior} 
to taking the large loop length limit \cite{ccn}. This gives a well-defined 
framework for computing the boundary reparameterization invariant pair 
correlation function which also preserves its {\em normalization} 
\cite{poltorus}: this is the key ingredient that enables a numerical 
prediction for the short distance potential
between heavy sources in the gauge theory
arising in fluctuations of the vacuum energy density.

Following Cohen et al \cite{cohen}, the tree correlation function for a 
pair of macroscopic string loops is represented as a path integral over 
embeddings and metrics on world-sheets of cylindrical topology terminating 
on fixed curves, $C_i$, $C_f$, within the spacetime $\D$, weighted by the 
locally supersymmetric action given in Eq.\ (\ref{eq:local}): 
\begin{equation}
< M(C_i) M(C_f)> ~=~
\half \sum_{(\beta,\alpha)\neq (1,1)} \C 
 \int_{[C_i,C_f][\beta,\alpha]}  
{{[d e][d\chi]}\over{{\rm Vol}(gauge)_{\partial M}}} 
  \int {{[dX][d\psi][d g][d\chi]}\over{{\rm Vol}(gauge)_M}} 
e^{-S_{SP}[X,\psi,g,\chi] - \mu_0 \int_{\cal M} \ds \sqg - S_{\rm ren.}} 
\quad .
\label{eq:pathf}
\end{equation}
As has been emphasized in section IIIA-B, we have taken all of the NN, DD, 
world-sheet fermions to satisfy identical boundary conditions at the 
end-points $\stwo$$=$$0$, $1$, for world-sheets of specified spin structure. 
We decouple bulk and boundary deformations of the world-sheet fields 
as in \cite{ccn}, imposing Dirichlet boundary conditions on all of the 
spatial embedding coordinates, $X^{\mu}$, $\mu$$=$$1$, 
$\cdots$, $9$. Then the boundary conditions on the matter fermions, 
$\psi^{\mu}$, are given by 
Eq.\ (\ref{eq:sigma2}). The distinction between the Wilson loop correlation
function and the ordinary annulus amplitude computed in section III comes 
from the inclusion of fluctuations in the world-sheet metric (einbein) on 
the boundaries of the annulus \cite{alvarez,cohen,ccn}. We gauge both 
boundary diffeomorphisms and local supersymmetry variations on the boundary. 
Anomalies of the measure under Weyl, and super-Weyl, transformations will, 
as before, be exponentiated as terms in the effective 
action proportional to the super-Liouville theory with boundary terms 
included \cite{polyakov,durhuus}. 
The quantization of the super-Liouville fields could be performed along
the lines of \cite{dhk}, and citations thereof, but we will be interested 
in the vacuum amplitude in the critical spacetime 
dimension where the super-Liouville theory decouples.

Consider the measure for einbeins. As was shown in \cite{ccn}, a 
reparameterization $\delta f(\sigma^1)$, tangential to the boundary
induces a non-trivial boundary Jacobian computed in \cite{cohen,ccn}. 
However, a Weyl rescaling of the einbein can always be absorbed in a shift 
in the Liouville field on the boundary. Consider the variation in 
the measure for einbeins under a local supersymmetry transformation.
From Eqs.\ (\ref{eq:loc})--(\ref{eq:weyl}) of appendix A, we see that 
the variation in the einbein under a supersymmetry transformation 
can always be absorbed in a rescaling of the super-Liouville fields, 
$(\phi,\zeta)$, on the boundary:
\begin{equation}
\delta_S e = 2 e^m_a (\delta_S e^a_m)
= -2 (\delta \xib) (\gamma^m \chi_m ), \quad \quad \delta_W e = 2 
\delta \lambda
\quad ,
\label{eq:suein}
\end{equation} 
where $\delta_S$, $\delta_W$, respectively denote the variation under 
local supersymmetry and Weyl transformations. Likewise, consider the 
variations
in the gravitino on the boundary. In superconformal gauge, setting
$\gamma^m \chi_m$$=$$0$, there are no independent variations of the 
gravitino on the boundary that have not already been accounted for in 
the analysis in section IIIB: a variation in $\chi$ on the boundary is a 
departure from superconformal gauge. The resulting super-Weyl anomaly is 
absorbed in the super-Liouville dynamics.
Thus, the sum over boundary deformations of the gravitino in 
Eq.\ (\ref{eq:pathf}) is pure gauge. We eliminate
the contributions to the measure from the zero modes of the
Neumann embedding coordinates in the expression for the
supersymmetric annulus derived in Eq.\ (\ref{eq:resamp}). The 
result for the pair correlation function of Wilson loops in the
T-dualized type I theory is a 
remarkably simple extension of the bosonic analysis 
given in \cite{ccn}:   
\begin{equation}
< M(C_i) M(C_f) > = 
 \int_0^{\infty} dl 
 e^{-S_{\rm saddle}[{\bar x}, {\hat{g}}]}
   ~ \eta(il )^{-12} 
 \sum_{(\beta,\alpha)\neq(1,1)} ~  
   \C ~ \Theta _{(\beta,\alpha)}^4 (0 , il ) 
\quad .
\label{eq:mesfermi}
\end{equation}
The annulus terminates on fixed curves, $C_i$, $C_f$, with fixed
spatial separation $R$ in some generic D-manifold, i.e., within 
the world-volume of a Dpbrane,
in the intersection of the world-volumes of
two or more Dbranes, or in the bulk space
transverse to some configuration of branes.
The path integral computes quantum fluctuations about a
saddle world-sheet configuration stretched between the
loops, $C_i$, $C_f$, and the saddle-point action can be computed
as in \cite{cohen,ccn}. In \cite{ccn} we focussed on
the simplest configurations of coplanar loops which can
capture the universal features of the large loop length 
limit which determines the short distance potential between
heavy point sources in the gauge theory. For such a configuration,
we showed in \cite{ccn} that the dominant contribution to the 
saddle action in
the small $R$, large $l$ limit takes the form,
$S_{\rm saddle}$$\sim$$ R^2 l/2\pi\alpha^{\prime}$,
independent of the shape or other geometrical characteristics
of the loops.

The reader may wonder if generalizations of this result with new
non-trivial fermionic degrees of freedom on the boundary are possible. 
The answer lies
in our understanding of the global structure of supermoduli space. The 
analysis given above is appropriate for the superconformal gauge fixed 
perturbative superstring in a Dbrane background. 
However, since brane dynamics is as yet a 
poorly understood subject in the wider context of String/M theory, it may 
be that the global structure of supermoduli space can play an interesting 
role in the full nonperturbative theory \cite{rabin}. 

\subsection{Generic Boundary Conditions on the Annulus}
\label{sec:motion}

Consider the supersymmetric annulus derived in Eq.\ (\ref{eq:resamp}) 
with boundaries on a pair of static and parallel Dpbranes. In this 
section, we sketch the modifications to Eqs.\ (\ref{eq:resamp}) 
for a configuration of rotated Dpbranes or, by an analytic continuation
of a Neumann coordinate, for a pair of Dpbranes in relative motion,
previously derived in \cite{bachas,polchinskibook}. It is 
convenient to complexify the coordinates, $X^{\mu}$, $\mu$$=$$1$,
$\cdots$, $8$, in pairs, decomposing a generic rotation into independent 
rotations in the four planes, $(1,2)$, $(3,4)$, $(5,6)$, $(7,8)$. The 
$(1,1)$ fermionic path integral no longer vanishes for generic boundary 
conditions on the fermions. For the generic rotation in 
all four planes, all four sectors of the fermionic path integral, 
$\beta,\alpha$$\in$$0,1$, contribute 
to the one-loop vacuum amplitude \cite{polchinskibook}. This case is 
discussed in appendix B. We will consider here
the simpler case of rotation in a single plane. The unknown phases in 
the annulus amplitude are determined by an extension of the infrared 
consistency conditions described in section IIIB for configurations 
of moving Dbranes. As a consistency check, we verify that we recover
the long distance velocity dependent potential between Dpbranes in relative 
motion, including the numerical coefficient previously computed in 
\cite{polchinskibook}.
This result will be adapted in section IV to obtain the supersymmetric 
pair correlation function of Wilson loops corresponding to closed 
world-lines of heavy sources in a gauge theory in relative slow motion.

Consider motion in the plane $X^0$$=$$iX^2$, $X^1$. The functional 
determinant for a complex scalar satisfying the V boundary conditions 
given in Eq.\ (\ref{eq:gbc}) can be obtained using zeta function 
regularization \cite{polchinskibook,ccn}. Likewise for the 
corresponding Weyl fermion; the fermionic functional determinant 
with spin structure $\alpha$ and boundary condition $\beta$ is simply 
the expression given in Eq.\ (\ref{eq:alpha}) with nonvanishing
argument for the Jacobi theta functions: $\Theta_{(\beta,\alpha)}(\nu,\lh)$, 
with $\nu$$=ul/2\pi$ \cite{bachas,polchinskibook}. Thus,
\begin{equation}
q^{E_0(\beta,u)} \prod_{m=1}^{\infty}
(1+ z~ q^{m- \mmi })(1+ z^{-1} q^{m- \mpl }) 
= \left [ 
{{ e^{u^2 l/2 \pi } \Theta _{(\beta,\alpha)} (ul/2\pi , 
\lh )} \over{ \eta(\lh) }} \right ] 
\quad ,
\label{eq:alphau}
\end{equation}
where the parameter $z$$=$$ e^{ i \pi (\alpha + ul/\pi )}$. The
vacuum energy of the Weyl fermion with the V boundary condition is
$E_0$$=$$\twofour$$+$$ \half ({{iu}\over{\pi}}$$+$${{\beta}\over{2}})^2$. 
Thus, the one-loop vacuum amplitude with boundaries on 
Dpbranes in relative motion takes 
the form \cite{bachas,dkps,polchinskibook}: 
\begin{equation}
\A (r,u) = {{1}\over{2}} V_{p} \int_0^{\infty} {{dl}\over{l}} 
(4 \pi^2 \alpha^{\prime}l )^{-\phalf} 
{{e^{-r^2 l/4\pi\alpha^{\prime}} 
\eta(\lh )^{-9}}\over{i \Theta_{11}(ul/2\pi,\lh )}}  
 \sum_{(\beta,\alpha)\neq(1,1)}  \C 
\Theta _{(\beta,\alpha)}^3 (0,\lh )\Theta_{(\beta,\alpha)} 
(ul/2\pi , \lh ) \quad .
\label{eq:resampu} 
\end{equation}
It is convenient to rescale $l$$\to$$2l$ in the final result.

The phases in the vacuum amplitude can be determined by infrared 
consistency conditions. As in section IIIB, we require the absence of
a tachyon and the vanishing of the static force between the branes. 
At long distances, we 
must recover at order $v^4$ the attractive $- 1/r^{7-p}$ potential 
between BPS sources required from matching to
the low energy type II supergravity theory. 
The long distance physics of the vacuum amplitude is dominated by the 
lowest lying closed string modes, exposed 
by taking the $l$$\to$$0$ limit of the expression in 
Eq.\ (\ref{eq:resampu}). Using standard identities for the Jacobi theta 
functions \cite{ww,polchinskibook} we can write:
\begin{eqnarray}
\A (r,u) =&& -{{1}\over{2}} V_{p} \int_0^{\infty} {{dl}\over{l}} 
(8 \pi^2 \alpha^{\prime} )^{-\phalf} 
{{ e^{-r^2 l/2\pi\alpha^{\prime} } l^{(6-p)/2} }
   \over{ \eta(\li )^{9} \Theta_{11}( \uar , \li ) }}  
 \sum_{(\beta,\alpha)\neq(1,1)} ~  \C ~ 
  \Theta _{(\alpha,\beta)}^3 (0 , \li ) 
~ \Theta_{(\alpha,\beta)} (\uar , \li )  
\nonumber 
\\ 
 =&& {{1}\over{2}} V_{p} (8 \pi^2 \alpha^{\prime} )^{-\phalf} 
\int_0^{\infty} dl e^{-r^2 l/2\pi\alpha^{\prime} } 
  {{ l^{(4-p)/2} q^{-1/2} }\over{2 \su }} 
\nonumber 
\\ 
&& \quad \times \left \{ q^0 (C_0^0 + C_0^1 ) + \qh [(2 \ctu + 6)
(C_0^0 - C_0^1 ) + 16 C_1^0 ~ {\rm Cos}(-iu) ] + O(q) \right \} 
\quad .
\label{eq:vacuinvu}
\end{eqnarray}
The absence of the closed string tachyon appearing at $O(q^{-1/2})$ 
requires $C_0^1$$=$$-C_0^0$. With this choice, a small $u$ expansion 
of the $O(q^0)$ terms gives:
\begin{equation}
C_0^0 ( 16 + 8 u^2 + {{8}\over{3}}u^4 +O(u^6)) + C_1^0 ( 16 + 8 u^2 + 
  {{2}\over{3}} u^4 + O(u^6) ) \quad . 
\label{eq:massless}
\end{equation}
The absence of a static force between the Dbranes requires that we set 
$C_1^0$$=$$-C_0^0$. As a consequence the leading contribution to the 
vacuum amplitude is $O(u^4)$--- at which order, 
spacetime supersymmetry is broken. The non-vanishing coefficient 
implies the existence of a long range velocity dependent potential 
between the Dbranes \cite{bachas,dkps}. The sign of the potential is 
determined once we fix the phase $C^0_0$. Defining the 
Minkowskian potential 
as in Eq.\ (\ref{eq:potentil}) and substituting 
for the phases in Eq.\ (\ref{eq:vacuinvu}) gives:
\begin{eqnarray}
V_{\rm long}(r,u) &&= - C_0^0 ~ 2 V_{p} (8 \pi^2 \alpha^{\prime} 
  )^{-\ponehalf} 
{{1}\over{2}} \int_0^{\infty} dl e^{-r^2 l/2\pi\alpha^{\prime} } 
  l^{(5-p)/2} ~{{ {\rm tanh} (u)}\over{2 i {\rm Sin} (-iu)}}
     [ 2 u^4 + O(u^6) ]  
\nonumber
\\
 &&= - u^4 C_0^0 ~  V_{p} (8 \pi^2 \alpha^{\prime} )^{-\ponehalf} 
           {{1}\over{r^{7-p}}} \Gamma ({{7-p}\over{2}}) 
              (2\pi \alpha^{\prime})^{(7-p)/2} + O(u^6) 
\quad . 
\label{eq:polongu}
\end{eqnarray} 
The potential is required to be {\em attractive} which implies 
$C_0^0$$=$$1$. Thus, we recover the potential between Dpbranes
including the numerical coefficient previously computed in
\cite{polchinskibook}:
\begin{equation}
V_{\rm long}(r,u)= - {{u^4}\over{r^{7-p}}} 
V_{p} 2^{2-2p} \pi^{(5-3p)/2} {\alpha^{\prime}}^{3-p} 
\Gamma ({{7-p}\over{2}}) + O(u^6) 
\quad .
\label{eq:potlu}
\end{equation} 
Thus, all three phases in Eq.\ (\ref{eq:resampu}) are determined,
$C_0^0$$=$$1$, $C_0^1$$=$$C_1^0$$=$$-1$, and we
have an unambiguous expression for the amplitude.

An extension of these arguments can be applied to more complicated
non-BPS brane configurations. Brane configurations which break one-half 
(one-quarter) of the spacetime 
supersymmetries can be distinguished by requiring that an order $v^2$ 
velocity dependent force is 
respectively absent (present). Similar arguments apply to configurations 
of mixed, intersecting, or rotated 
Dbranes: from the low energy correspondence to supergravity, we infer the 
qualitative form of the long distance potential. This is then applied 
as an infrared consistency 
condition on the unknown phases in the supersymmetric annulus.

\section{Short Distance Potential Between Heavy Sources}
\label{sec:potential}

In an earlier work \cite{ccn}, we showed that there exists a short 
distance interaction between heavy sources in a gauge theory
traversing fixed spacetime paths in some generic 
background of the bosonic string. The potential arises in fluctuations in the 
vacuum energy density. The same phenomenon can be observed in the 
supersymmetric T-dualized type I string theory. We will obtain
in this section an expression for the short distance potential 
between heavy sources in a supersymmetric gauge theory. Our results
are derived in a systematic small velocity short distance double expansion,
following an analogous treatment of the short distance
potential in bosonic string theory in \cite{ccn}. The potential is 
extracted from the supersymmetric pair correlation function of 
Wilson loops, in the limit of large loop lengths and small spatial 
separations.

We begin with the pair correlation function of Wilson
loops corresponding to world-lines of heavy sources in relative collinear 
motion with nonrelativistic velocity $v$ in the direction $X^1$.
For coplanar loops, this mimics straight line trajectories in the 
Euclideanized $X^0$$=iX^2$, $X^1$ plane: $r^2$$=$$R^2$$+$$v^2\tau^2$,
for small separations $r$. The modifications to the expression given in 
Eq.\ (\ref{eq:mesfermi}) for these boundary conditions are straightforward,
following the results of section IIID. In the 
large loop length limit, we have:
\begin{equation}
< M(C_i) M(C_f) > = 
 \int_0^{\infty} dl {{ e^{-r^2 l/2\pi\alpha^{\prime}}
   \eta(il )^{-9} }\over{i \Theta_{11}(ul/\pi,il)}} 
 \sum_{(\beta,\alpha)\neq(1,1)} ~  
   C_{\alpha}^{\beta} ~ \Theta _{(\beta,\alpha)}^3 (0 , il ) 
    \Theta _{(\beta,\alpha)} (ul/\pi , il ) 
\quad .
\label{eq:mesfermiu}
\end{equation}
The short distance dynamics is dominated by 
the lowest lying modes in the open string spectrum. We define the 
Minkowskian potential \cite{ccn}: 
\begin{equation}
< M(C_i) M(C_f) > = -i \lim_{T\to\infty} \int_{-T}^{+T}
    d\tau V_{\rm eff.}[r(\tau),u]
\quad .
\label{eq:potentii}
\end{equation}
The short distance regime is exposed by expanding the integrand in
Eq.\ (\ref{eq:mesfermiu}) in powers of $q$$=$$e^{-2\pi l}$.
Substituting for 
the phases, we can express the Minkowskian potential at short 
distance in a small $q$ expansion as in \cite{ccn}:
\begin{eqnarray}
V_{\rm eff.}(r,u) =&& 2 (8 \pi^2 \alpha^{\prime} )^{-1/2} 
\int_0^{\infty} dl ~ {\rm tanh} (u) l^{1/2}  
 {{e^{-r^2 l/2\pi\alpha^{\prime}} \eta(il)^{-9}}\over{\Theta_{11}(ul/\pi,il)}}  
   \sum_{(\beta,\alpha)\neq(1,1)}  C_{\alpha}^{\beta} \Theta_{(\beta,\alpha)}^3 
     (0,il) \Theta_{(\beta,\alpha)} (ul/\pi , il ) 
\nonumber \\
 =&& - (8 \pi^2 \alpha^{\prime} )^{-1/2} \int_0^{\infty} dl ~ 
 e^{-r^2 l/2\pi\alpha^{\prime}} 
{{l^{1/2} {\rm tanh} (u) }\over{ \qh {\rm Sin} (ul)}}   
\nonumber \\
&& \quad \times \left [ (1+2 \qh )^3 (1+2\qh \cul ) - (1-2\qh)^3(1-2\qh\cul) -
16 \qh \culh + O(q) \right ] 
\quad .
\label{eq:potsh}
\end{eqnarray}
The leading non-vanishing terms in this expression, due to massless exchange,
contribute at order $q^0$:
\begin{equation}
V(r,u) = - (8 \pi^2 \alpha^{\prime} )^{-1/2} \int_0^{\infty} dl ~ 
 e^{-r^2 l/2\pi\alpha^{\prime}} 
{{l^{1/2} {\rm tanh} (u) }\over{  {\rm Sin} (ul)}}   
 \left [ 12 + 4 {\rm Cos} (2ul) - 16 {\rm Cos}(ul) \right ]
\quad .
\label{eq:potmsh}
\end{equation}
We have assumed small velocities $v$$=$${\rm tanh}(u)$$\simeq$$u$. We now 
perform a resummation of the integrand in the variables $r$,$u$, as in \cite{ccn}. 
The regime of validity is determined by the behavior of the cosecant 
function. We perform a Taylor expansion in the first half-period of its
argument, $0$$\le$$ul$$<$$\pi$. For sufficiently small $u$ values 
the oscillations in the integrand are increasingly
rapid, smearing out the integral \cite{polchinskibook}. As in the analysis of
the vacuum amplitude for the bosonic string in \cite{ccn}, we note that 
the contribution from the integration domain $ul$$\ge$$\pi$ can
always be bounded, or evaluated by numerical integration, as a 
self-consistency check on the approximation. This check provides an upper 
limit, $u_+$, on the permissible velocities. With this
restriction, the contribution from the domain $l$$>$$\pi/u_+$ can be
dropped and we suppress it in what follows. 
The potential takes the form:
\begin{equation}
V(r,u) = - (8 \pi^2 \alpha^{\prime} )^{-1/2} 
\int_0^{\pi/u_+} dl ~ e^{-r^2 l/2\pi\alpha^{\prime}} 
l^{-1/2} {\rm tanh}(u)/u    
\left [ \sum_{k=1}^{\infty} C_k (ul)^{2k} +
\sum_{k=1}^{\infty} \sum_{m=1}^{\infty} C_{k,m} (ul)^{2(k+m)}
\right ]
\quad ,
\label{eq:expand} 
\end{equation}
where the coefficients in the summation are given by:
\begin{eqnarray}
C_k =&& {{ 4 (-1)^k (2^{2k} - 4 )}\over{(2k)!}} 
\nonumber \\
C_{k,m} =&& {{ 8 (-1)^k (2^{2m-1} -1)}\over{(2k)! (2m)!}} ~ 
 |B_{2m}| ~(2^{2k} - 4)
\quad .
\label{eq:coeff} 
\end{eqnarray}
The $B_{2m}$ are the Bernoulli numbers. Note that the $k$$=$$1$ term vanishes
in both sums and the leading velocity dependence of the amplitude is $O(u^4)$.  
Integrating over $l$ gives a systematic expansion for
the potential in powers of $u^{2}/r^{4}$. As in \cite{ccn}, we identify a
dimensionless scaling variable, $z$$=$$r^2_{\rm min.}/r^2$,
where $r_{\rm min.}^2$$=$$2 \pi \alpha^{\prime} u$. The
velocity dependent corrections to the potential between heavy sources in the
supersymmetric gauge theory are succinctly
expressed as a convergent power series in the {\em single} 
dimensionless variable $z$:
\begin{eqnarray}
V(r,u) =&& - (8 \pi^2 \alpha^{\prime} )^{-1/2} {\rm tanh}(u)/u    
\cdot r^{-1} (2\pi\alpha^{\prime})^{1/2} [
\sum_{k=1}^{\infty} C_k z^{2k} \gamma (2k +1/2,\pi/z)
\nonumber \\
&&\quad\quad + \sum_{k=1}^{\infty} \sum_{m=1}^{\infty} C_{k,m} z^{2(k+m)}
\gamma (2(k+m) + 1/2,\pi/z)
 ] 
\quad .
\label{eq:vexpn}
\end{eqnarray}
Note that the potential takes the form of a scale invariant
$1/r$ fall-off, contributed by the bosonic degrees of freedom \cite{ccn},
multiplicatively corrected by a convergent power series in $z$:
\begin{equation}
V (r,u) = - 
{{{\rm tanh}(u)/u}\over{\Gamma(\half)}} {{1}\over{r}} 
\left [ z^4 \gamma({{9}\over{2}},\pi/z) + O(z^6) \right ]
\quad ,
\label{eq:dimles}
\end{equation}
indicative of its origin in fluctuations of the vacuum energy density.
Recall that the regime of validity for the double expansion in small
velocities and short distances is $z$$<$$1$, $u$$<$$u_+$. 
The scale factor $z$ determines the magnitude of the velocity dependent
corrections and, therefore, the accuracy of the expansion. For a given
accuracy, with fixed $z$ value, we can probe arbitrarily short distances
$r$ by simultaneously adjusting the velocity.
The power series corrections in the superstring are qualitatively similar, 
but much simpler than the analogous series in the bosonic string theory 
\cite{ccn}: the analogous result in a nonsupersymmetric gauge theory
receives corrections in the variables 
$z^2$, $uz/\pi$, and $u^2$. The leading term in the potential between
heavy sources in a supersymmetric gauge theory is therefore $O(u^4/r^9)$:
\begin{equation}
V(r,u) = - {{u^4}\over{r^{9}}}  
2^{4} \pi^{7/2} {\alpha^{\prime}}^{4}
\Gamma({{9}\over{2}})
+ O(u^6)
\quad .
\label{eq:vshort}
\end{equation}

\section{Conclusions}
\label{sec:conc}

We have given a derivation from first principles of both the normalization
and the phase of the supersymmetric annulus in the generic flat D-manifold
background of the type I and type II string theories. The normalization of 
the string path integral is determined by its symmetries \cite{poltorus}. 
As a consequence, one can extract numerical predictions from one-loop string 
amplitudes, free from any dependence on the string coupling constant. We have
shown in this paper
that phase ambiguities in the fermionic string path integral can be 
eliminated by the imposition of simple, and universal, infrared consistency 
conditions on {\em qualitative} features of the long distance physics, by 
matching to an appropriate supergravity theory. This prescription gives the 
same result as the usual GSO projection in the superstring, but has the 
hope of wider applicability to generic backgrounds of String/M theory. We 
note that we have emphasized the BPS conditions over supersymmetry. This 
is in keeping with the broader goal of understanding the self-Duality of 
String/M theory in generic backgrounds for the Ramond-Ramond antisymmetric 
tensor fields associated with Dbrane charges \cite{polchinskibook,witten}. 
The preliminary results given here need to be explored in a wider context,
extended to an understanding of the consistency conditions that suffice 
to ensure tadpole cancellation \cite{polwit,horwit,gimpol} in generic 
backgrounds of String/M theory.

Extending our earlier results for the bosonic string \cite{ccn}, we have 
shown that heavy point sources in a supersymmetric gauge theory in slow 
relative motion have an attractive, and velocity dependent, interaction 
at short distances. The potential can be expressed as a convergent power 
series in the single dimensionless variable $z$$=$$r_{\rm min.}^2/r^2$, 
where $r_{\rm min.}^2$$=$$2 \pi \alpha^{\prime} u$ is the minimum distance 
probed in this approximation, valid for small velocities and short distances
in the regime $2\pi\alpha^{\prime} u $$<<$$r^2$$<<$$2 \pi\alpha^{\prime}$.
It would be gratifying if this result could be exploited as a window into 
the short distance physics of String/M theory.

\vspace{0.3in}
\noindent{\bf Acknowledgments}

This work is supported in part by the National Science Foundation grant
NSF-PHY-97-22394.

\appendix
\section{Spinor Conventions and Local Symmetries}
\label{app:spinor}

In this appendix we establish our conventions, simultaneously deriving the 
Brink-Di Vecchia-Howe-Deser-Zumino action for the fermionic string 
with Minkowskian signature metric \cite{dzbdh}. The classical action is 
invariant under both 
reparameterizations and $N$$=$$1$ world-sheet supersymmetry transformations.
The Euclidean action used in the string path integral is obtained by an 
analytic continuation.
We consider free massless spinor fields in a two dimensional 
Minkowskian space parameterized $(t,x)$ with metric:
\begin{equation}
\{ \gmu , \gnu \} = 2 \eta^{\mu\nu } , 
 \quad\quad \joz = \ifour [\gzero ,\gone ]= \itwo \gamma  
\quad .
\label{eq:cliff}
\end{equation}
$\joz$ is the sole generator of Lorentz transformations in two dimensions and 
the matrix $\gamma$ projects onto spinors of definite chirality. We choose a 
representation with real gamma matrices. Thus,
\begin{equation}
\label{eq:generators}
\gamma^0 = \left( \begin{array}{rr}
                 0 & 1 \cr
                 -1 & 0 \end{array} 
\right), ~~~~~
\gone  =  \left( \begin{array}{rr}
                     0 & 1 \\
                     1  & 0 \end{array} \right), ~~~~
\gamma = \gzero\gone = \left( \begin{array}{rr}
                     1 & 0 \\
                     0 & -1 \end{array} \right).
\nonumber
\end{equation}
Fermion bilinears that transform as scalars under Lorentz transformations
are obtained by identifying a matrix $\beta$ such that:
\begin{equation}
\beta (\dl)^\dagger \beta = \dli , \quad \beta (\jmunu)^\dagger \beta =\jmunu , 
\quad\quad {\rm where} ~ \dl \equiv e^{\itwo (\wmunu \jmunu)}  
\quad .
\label{eq:dlorbe}
\end{equation}
Note that with this choice of gamma matrices the Lorentz generator
acts non-unitarily in the spinor representation: $(\dl)^\dagger$$\neq$$\dli$. 
Thus, $\beta$ must be chosen to satisfy the conditions: 
\begin{equation}
\beta (\gmu)^\dagger \beta=-\gmu , 
\quad\quad \beta (\jmunu)^\dagger \beta =\jmunu  \quad ,
\label{eq:beta}
\end{equation}
with solution $\beta$$=$$ i\gzero$. It is easy to verify that the 
fermion bilinear:
\begin{equation}
\psipl \beta \psi \to \psipl (\dl)^\dagger \beta \dl \psi = 
  \psipl \beta (\dli \dl ) \psi=\psipl \beta\psi
\equiv \psib\psi
\quad ,
\label{eq:fbilin}
\end{equation}
is Lorentz invariant. Defining components, 
$(\psi_\mu)^T$$=(\psipl_\mu,\psimi_\mu)$, 
the free fermion Lagrangian on flat world-sheets can be written 
in component form:
\begin{equation}
\L = - \psibmu \gm \pm \psi^-_\mu = i (\psi^{-\mu})^{*} (\pzero +\pone) \psimi_\mu +
i (\psi^{+\mu})^* (\pzero -\pone) \psipl_\mu \quad .
\label{eq:lag}
\end{equation}
Notice that, with the conventions above, charge conjugation is defined by the 
Majorana condition, $\zeta^*$$=$$\gamma\zeta$, with spinors
$\zeta$ and $\gamma(\zeta)^*$
transforming identically under Lorentz transformations:
\begin{equation}
\gamma (\gmu) \gamma^{-1} = 
\gamma (\gmu ) \gamma = - (\gmu)^* , \quad \quad \gamma ( \jmunu ) \gamma^{-1} = -(\jmunu)^*
\quad .
\label{eq:dlor}
\end{equation}
We can therefore choose the component fermions, $\zeta^+$, $\zeta^-$, to be real.

Analytically continuing to world-sheets with Euclidean signature, we can set
$\sone$$=$$it$, $\stwo$$=$$x$. Thus, we replace, $\pzero$$\to$$i\pone$, 
$\pone$$\to$$\ptwo$, in Eq.\ (\ref{eq:lag}) to obtain the Euclidean Lagrangian in 
component form: 
\begin{equation}
\Le =  \psi^{-\mu} (\pone - i \ptwo) \psimi_\mu +
   \psi^{+\mu} (\pone + i \ptwo ) \psipl_\mu \quad .
\label{eq:lage}
\end{equation}
It is easy to verify that Eq.\ (\ref{eq:lage}) can be recovered from the Lagrangian, 
$\Le$$=$$\psib^\mu \ga\pa\psi_\mu$, $\psib$$\equiv$$(\psi)^T\gone$ with the
following choice of gamma matrices with Euclidean metric:
\begin{equation}
\label{eq:Egenerators}
\gone  =  \left( \begin{array}{rr}
                     0  & i \\
                     -i & 0  \end{array} \right),
                     ~~~~
\gtwo  =  \left( \begin{array}{rr}
                     0 & 1 \\
                     1 & 0 \end{array} \right), ~~~~
\gamma^5 = \gamma = -i \gone \gtwo = \left( \begin{array}{rr}
                     1 & 0 \\
                     0 & -1  \end{array} \right).
\nonumber
\end{equation}
From the equation of motion, it is clear that the component fermions, 
$\psi^+$, $\psi^-$, transform, respectively, as left-handed, and 
right-handed, spinors on the world-sheet.
We remark that our spinor conventions correspond to those used in the 
text \cite{polchinskibook}. The reader can verify the identities:
\begin{equation}
\xib \chi = (\xi)^T \gone \chi = i(\chipl\ximi - \ximi \chipl) = \chib\xi
, \quad \xib\ga\chi=- \chib\ga\xi , \quad \xib\ga\gb\chi=\chib\gb\ga\xi \quad . 
\label{eq:ids}
\end{equation}

The free fermion Lagrangian can be extended to a two-dimensional action 
invariant under a local $N$$=$$1$ supersymmetry following \cite{dzbdh}. 
We must be careful to retain any boundary terms resulting from an 
integration by parts since our interest is in the classical action for
world-sheets with boundary. Appending $d$ free fermions to the bosonic 
string world sheet gives the free field action:
\begin{equation}
S_0 =  \quaft \int d^2 \sigma [ \partial^a \Xmu \pb X_{\mu} + 
   \ap ( \psi^{-\mu} (\pone - i \ptwo) \psimi_\mu +
   \psi^{+\mu} (\pone + i \ptwo ) \psipl_\mu ) ] \quad .
\label{eq:flat}
\end{equation}
A variation of the free field action under the global supersymmetry 
transformation:
\begin{equation}
\delta \Xmu = \sqap (\xib \psi^{\mu}) , \quad  
\sqap (\delta \psi^{\mu}) = \half (\pa \Xmu ) \ga \xi , \quad 
\sqap (\delta \psib^{\mu}) = - \half \xib \ga (\pa \Xmu)  \quad , 
\label{eq:glob}
\end{equation}
results in the variation, 
\begin{equation}
 \2p (\delta {\cal L_0}) = - \sqap 
\left \{ \pa [ \half (\pb X_{\mu})(\xib \ga\gb\psi^\mu) ] 
 + (\pa \xib ) (\pb X_{\mu}) (\xib \ga\gb\psi^\mu) \right \} 
\quad .
\label{eq:var}
\end{equation}
Note that the second term vanishes for a covariantly constant spinor 
supersymmetry parameter, $\xi$. We can identify the Noether current, 
$J^a$$=$$(\pb \Xmu )(\gb\ga\psi_{\mu})$, and introduce a fermionic source term
in the Lagrangian: $\chib_aJ^a$, where $\pa \xi$$\equiv$$-(\delta \chi)$
\cite{dzbdh}. The resulting variations close upon inclusion of a new term 
quartic in the fermions. The result is the Deser-Zumino-Brink-De Vecchia-Howe 
action for the fermionic string \cite{dzbdh}. Introducing the world-sheet 
metric, $g_{mn}$$=$$e_m^a e_{na}$, we can write the action in covariant form:
\begin{equation}
S =  \halft \int d^2 \sigma \sqg [ \half g^{mn}\partial_m \Xmu \pn X_{\mu} + 
   \half \ap \psib^\mu \gm \partial_m \psi_\mu + \sqap (\chib_a \gm\ga \psi^{\mu})
  (\partial_m X_{\mu}) + \apfour (\chib_a\gb\ga\psi^{\mu})(\chib_b\psi_{\mu}) ]
\quad ,
\label{eq:local }
\end{equation}
invariant under the local supersymmetry transformations:
\begin{eqnarray}
\delta \Xmu &&= \sqap (\delta \xib) \psi^{\mu}  
\nonumber \\
\sqap (\delta \psi^{\mu}) &&= \half (\pa \Xmu ) \ga \delta \xi 
+ \half \sqap (\psi^{\mu} \chi_m) \gamma^m \delta \xi
\nonumber \\
\delta \chi_m &&= - D_m (\delta \xi) 
\nonumber \\
\delta e_m^a &&= - (\delta \xib) \ga \chi_m \quad . 
\label{eq:loc}
\end{eqnarray}
The action is invariant under both reparameterizations and supersymmetry
transformations on the world-sheet. The constraints that arise from making 
these compatible on the boundary will be discussed in Section II. Let us briefly 
recall the local invariances of the world-sheet action in the bulk. Under a 
diffeomorphism of the world-sheet coordinates parameterized by $\delv^m$, we have:
\begin{eqnarray}
\delta e_m^a &&=  \delv^n (\pn \ema) + \ena (\partial_m \delv^n) 
\nonumber \\
\delta \chi_m &&= \delv^n (\pn \chi_m) + \chi_n (\partial_m \delv^n) 
\nonumber \\
\delta \Xmu &&= \delv^n(\pn \Xmu) 
\nonumber \\
\delta \psi^{\mu} &&= \delv^n (\pn\psi^{\mu}) 
\quad . 
\label{eq:diff}
\end{eqnarray}
In the tangent space at any point of the manifold, we can 
perform independent Lorentz rotations, $\delta \omega$:
\begin{eqnarray}
\delta e_m^a &&=  (\delta \omega ) \eab \emb 
\nonumber \\
\delta \chi_m &&= \half (\delta\omega ) \gfive \chi_m  
\nonumber \\
\delta \psi^{\mu} &&= \half (\delta\omega ) \gfive\psi^{\mu}  
\quad , 
\label{eq:lor}
\end{eqnarray}
which leave world-sheet scalars invariant. The scalars are also
invariant under rescalings of the world-sheet metric and 
gravitino. A Weyl rescaling of the metric, $\delta\Lambda$, 
induces the variations:
\begin{eqnarray}
\delta e_m^a &&= ( \delta \Lambda )\ema
\nonumber \\
\delta \chi_m &&= \half (\delta \Lambda) \chi_m 
\nonumber \\
\delta \psi^{\mu} &&= - \half (\delta \Lambda) \psi^{\mu} 
\quad . 
\label{eq:weyl}
\end{eqnarray}
The action is also invariant under the fermionic rescaling, 
$\delta\zeta$, 
of the world-sheet gravitino, $\delta \chi_m$$=$$ (\delta \zeta)\gamma_m$, 
known as a super-Weyl transformation, which leaves all other 
world-sheet fields fixed. 
Note that the total number of bosonic and fermionic gauge
parameters are equal: $(\delta \lambda,\delv^n,\delta\omega)$ and
$(\delta \xi , \delta \zeta)$ contain four parameters each. They
can be used to locally fix the world-sheet metric, $\ema$,
and gravitino, $\chi_m$, to their values in the superconformal 
gauge \cite{howe}.  

\section{Constant Spinors and the Path Integral 
in the $(\beta,\alpha) $$=$$ (1,1)$ Sector}
\label{app:supermoduli}

For the Dbrane backgrounds studied in this paper, with either parallel
and static branes, or a relative rotation in a single plane, the 
contribution to the vacuum amplitude from the path integral with
$(\beta,\alpha)$$=$$(1,1)$ vanishes. As mentioned in the text, this is 
always true unless one considers rotations in all four transverse 
planes: $(1,2)$, $(3,4)$, $(5,6)$, and $(7,8)$. By an analytic continuation,
this implies that the contribution from the $(1,1)$ sector to the potential 
between point sources derived in section IV vanishes unless we consider a 
general motion with velocity components in {\em at least four} spatial 
directions. This can be compared with the discussion of a pair of D4branes 
in relative motion in four transverse Dirichlet directions given in 
\cite{polchinskibook}. From the point of view of the path integral, the 
vanishing contribution is due to a Grassmann integration over a constant mode 
which is absent in the action. For motions in four transverse planes, the 
constant mode is absent for four of the Weyl fermions on the world-sheet. We 
will verify in this appendix that the Grassmann integration over the 
constant mode of the remaining Weyl fermion, $(0,9)$, can be saturated by
an insertion in the path integral that comes from the supermodulus in the 
$(1,1)$ sector. 

We complexify the component fermions as in Eq.\ (\ref{eq:complex}) giving a 
total of five independent Weyl fermions on the world-sheet upon imposing the
open string boundary conditions. Consider separating the constant mode, 
$\psi_0^{\pm i}$, $i$$=$$1$, $\cdots$, $5$, with $\psi^{\pm i} $$=$$ \psi_0^{\pm i} 
$$+$$ (\psi^{\pm i})^{\prime}$. We have,
\begin{equation}
\label{eq:zeromode}
\int [d\psi] e^{-S(\psi)} \rightarrow \int \prod d\psi_0 \int[d\psi^\prime]
 e^{-S(\psi^\prime)} \quad .
\end{equation}
Since the integrand is independent of the constant mode, the Grassmann integral 
will vanish. We will now show that the fermionic Jacobian, $J_f$, defined in 
Eq.\ (\ref{eq:gravmes}) has a nontrivial dependence on the supermodulus in the
$(1,1)$ sector. This term can be exponentiated as a correction to the effective 
action for the fermions with the consequence that the Grassmann integration over 
{\em one} pair of constant fermion modes no longer vanishes. If these are the only 
fermionic zero-modes present, one obtains a non-zero contribution to the vacuum 
amplitude from the $(1,1)$ sector.

As discussed earlier, in the $(1,1)$ sector the cylinder has both a 
supermodulus and a superconformal Killing spinor, both of which are simply 
constant spinors on the world-sheet. We consider variations of the gravitino
that preserve the gamma tracelessness condition, parallel to the gauge slice. 
Denoting the supermodulus as $\nu$, a constant two component spinor, we 
can write,
\begin{equation}
\label{eq:supermodulus}
\chi_1 = \nu ,  \quad \quad \chi_2 = 
i \gamma^5 \nu
\quad .
\end{equation}
A traceless variation of the gravitino can be decomposed as
$\delta\chi_m = -D_m\xi^\prime + \chi_{m,\alpha} d\nu^\alpha$, where $\alpha$ 
labels the two spin-components of $\nu$. Substituting in Eq.\ (\ref{eq:normxi})
gives:
\begin{eqnarray}
\label{eq:delchi}
|\delta\chi_m|^2 = 
  -i\int d^2\sigma \sqrt{g}(\delta\bar{\xi}^\prime  d\nu^\alpha ) 
     \left( \begin{array}{cc}
      -D_mD^m & D^m \chi_{m, \beta} \\
      \bar{\chi}_{m,\alpha}D^m & \bar{\chi}_{m,\alpha} \chi^m_{,\beta} \end
{array} \right) \left( \begin{array}{c} \delta\xi^\prime \\
d \nu^\beta \end{array}
\right)
\quad .
\end{eqnarray}
Substituting in Eq.\ (\ref{eq:normg}),
\begin{equation}
1 = \int [d\delchi] e^{-|\delchi|^2/2} = J_f (\hat{g}) 
 \int [d\delxi_0] \int [d\delxi^{\prime}] \int d\nu e^{-|\delchi|^2/2} \quad .
\label{eq:normgg} 
\end{equation}
Including the contributions from the Grassmann integrations over the super
conformal Killing spinor and the supermodulus given in Eqs.\ (\ref{eq:normxisck}), 
(\ref{eq:normxism}), and substituting for the Jacobian matrix in 
Eq.\ (\ref{eq:delchi}), gives the result:
\begin{equation}
\label{eq:fermjacobian}
J_f(\hat{g}) = [\det(P_{1/2}^{\dagger}P_{1/2})]^{-1/2}(1/l^2 + 1)^{-1}
\quad .
\end{equation}
Thus, the fermionic path integral takes the form:
\begin{eqnarray}
\int_{[1,1]} \frac{[d\delta\chi_m]}
  {\rm{Vol(sWeyl \times sDiff)}}
 && e^{ \frac{1}{4\pi} 
 \int d^2\sigma \sqrt{g}(\bar{\chi}^m \psi^\mu)(\bar{\chi}_m\psi_\mu) } 
\nonumber \\
&&= \int_{[1,1]} d\nu [{\rm det}^{\prime} (P_{1/2}^{\dagger}P_{1/2})]^{-1/2}(1/l^2 + 1)^{-1}
 e^{(1/l^2+1)\frac{1}{2\pi}\int d^2\sigma \sqrt{g}(\nu^+\nu^-\psi^{+\mu
}\psi^-_\mu)}
\quad ,
\label{eq:substitution}
\end{eqnarray}
where we use component form for the fermions in the action.
Integrating over $\nu$, we are left with the following insertion in the path 
integral for the matter fermions:
\begin{equation}
\label{eq:insertion}
[{\rm det}^{\prime}(P_{1/2}^{\dagger}P_{1/2})]^{-1/2}\frac{1}{2\pi}\int d^2\sigma \sqrt{g
}(\psi^{+\mu}\psi^-_\mu)
\end{equation}
As mentioned in the text, the functional determinant, 
$[\det^{\prime}(P_{1/2}^{\dagger}P_{1/2})]^{-1/2}$, precisely cancels the 
contribution to the amplitude from the non-constant modes of one pair of 
component fermions. We choose these to be the $(0,9)$ pair. Complexify as 
described above. Summing on $i$$=$$1$, $\cdots$, $5$, the insertion takes 
the form,
\begin{equation}
\label{eq:zeromodes}
\frac{1}{2\pi}\int d^2\sigma \sqrt{g}(\psi^{+i }\psi^{-i}) 
= \frac{l}{2\pi}\psi^{+5}_0\psi^{-5}_{0} + \ldots
\quad,
\end{equation}
where the $\cdots$ denote the dependence on the remaining constant modes,
if present. Thus, precisely one pair of constant modes is saturated by
the insertion. If no additional constant modes are present, the resulting 
path integral gives a non-vanishing result. The fermionic oscillator 
contributions in the $(1,1)$ sector are computed by zeta function 
regularization as in Eq.\ (\ref{eq:alphau}). As an example, consider the 
relative motion of a pair of Dpbranes in the directions
$(X^1,X^3,X^5,X^7)$, with $v_i$$=$${\rm tanh}(u_i)$, $i$$=$$1$, $\cdots$, 
$4$. Then the contribution to the annulus from the $(1,1)$ sector takes 
the form:
\begin{equation}
\A_{[1,1]} (r,u) = {{1}\over{2}} V_{p} \int_0^{\infty} {{dl}\over{l}} 
(4 \pi^2 \alpha^{\prime}l )^{-\phalf} 
e^{-r^2 l/4\pi\alpha^{\prime}} 
\prod_{i=1}^4 {{\Theta_{(1,1)} (u_i l/2\pi, \lh )} 
\over{ i \Theta_{11}(u_i l/2\pi,\lh )}} 
\quad ,
\label{eq:resampuoneone} 
\end{equation}
previously derived in \cite{polchinskibook}.

\end{document}